\documentclass[pre,aps,floats,floatfix,superscriptaddress,nofootinbib,showpacs]{revtex4}
\usepackage[english]{babel}
\usepackage{amsmath,amssymb,amsfonts}
\usepackage[dvips]{graphicx}
\usepackage{amsmath}
\usepackage{fancyhdr}

\usepackage{bm}
\newcommand{\be}{\begin{equation}}
\newcommand{\ee}{\end{equation}}

\begin{document}

%%%%%%%%%%%%%%%%%%%% New commands %%%%%%%%%%%%%%%%%%%%%%%%%%%%%%%%%%%%
\newcommand{\sigmagorro}{\hat{\boldsymbol{\sigma}}}
%%%%%%%%%%%%%%%%%%%%%%%%%%%%%%%%%%%%%%%%%%%%%%%%%%%%%%%%%%%%%%%%%%%%%%
\title{Dynamics of Annihilation II: Fluctuations of Global Quantities}

\author{Pablo Maynar}
\affiliation{Laboratoire de Physique Th\'eorique (CNRS
  UMR 8627), B\^atiment 210, Universit\'e Paris-Sud, 91405 Orsay cedex,
  France}
\affiliation{F\'{\i}sica Te\'{o}rica, Universidad de Sevilla,
Apartado de Correos 1065, E-41080, Sevilla, Spain}
\author{Mar\'ia Isabel Garc\'ia de Soria}
\affiliation{Universit\'e Paris-Sud, LPTMS, UMR 8626, Orsay Cedex, F-91405 and
CNRS, Orsay, F-91405}
\author{Gr\'egory Schehr}
\affiliation{Laboratoire de Physique Th\'eorique (CNRS
  UMR 8627), B\^atiment 210, Universit\'e Paris-Sud, 91405 Orsay cedex, France}
\author{Alain Barrat}
\affiliation{Laboratoire de Physique Th\'eorique (CNRS
  UMR 8627), B\^atiment 210, Universit\'e Paris-Sud, 91405 Orsay cedex, France}
\author{Emmanuel Trizac}
\affiliation{Universit\'e Paris-Sud, LPTMS, UMR 8626, Orsay Cedex, F-91405 and
CNRS, Orsay, F-91405}

\date{\today }

\begin{abstract}

We develop a theory for fluctuations and correlations in a gas evolving under 
ballistic annihilation dynamics. Starting from the hierarchy of equations
governing the evolution of microscopic densities in phase space, we subsequently
restrict to a regime of spatial homogeneity, and obtain explicit 
predictions for the fluctuations and time correlation of the
total number of particles, total linear momentum and
total kinetic energy. Cross-correlations between these quantities are worked
out as well.
These predictions are successfully tested against 
Molecular Dynamics and Monte-Carlo simulations.
This provides strong support for the theoretical approach
developed, including the hydrodynamic treatment of the spectrum of the
linearized Boltzmann operator. This article is a companion paper to 
Ref. \cite{companion} and makes use of the spectral analysis reported
there.
\end{abstract}
\maketitle

\section{Introduction}

Systems where particles may react, change chemical or physical nature and ultimately 
disappear, model a rich variety of phenomena and provide
prominent situations to develop and test the foundations of non-equilibrium
statistical mechanics (see e.g. \cite{OZ78,tw83,rk83,bl88,l03} 
and references therein). 
When reactions are controlled ballistically, the system
can be modeled by an assembly of hard spheres or disks which annihilate with
probability $p$ or collide elastically with probability $1-p$ everytime two
particles meet each other \cite{cdt04s}. 
Within the framework of this probabilistic ballistic annihilation (PBA) model,
most of the work carried out up to now has focused on the kinetic equations
for the one-body distribution function and the information following from them
\cite{cdt04s,ks01,t02,ptd02,llf06}. In particular, the hydrodynamic equations, with
explicit expressions for the transport coefficients, have been derived by
using a generalization of the Chapman-Enskog expansion \cite{cdt04}. Our
companion paper \cite{companion}, where 
we have established the hydrodynamic description in the
context of the eigenfunctions and eigenvalues of the linearized Boltzmann
collision operator, falls in this vein. 
Within this formalism, the conditions in which the
hydrodynamic description is expected to apply are somehow more transparent,
and can be expressed in terms of some properties of the linearized Boltzmann
collision operator. 

In the present paper, the goal is to go beyond the study of one body quantities:
the focus is on fluctuations and
correlations. To this end, 
we use the tools and ideas developed in the context of the
linearized Boltzmann equation. The dynamical behavior of the correlations
in the dilute limit can indeed be obtained in terms of the 
linearized Boltzmann collision operator.  
The study of correlations in the PBA model 
allows to go beyond the description at the level of average values,
and to characterize how global magnitudes (such as the total number
of particles or the total energy) fluctuate around their 
average. It has already been shown in 
other classes of dissipative systems, such as in granular systems, how
the knowledge of fluctuations is relevant in order to understand the behavior
of the system when vortices or clusters develop \cite{nebo97,bmr98}, or even 
in simpler situations where the system is still homogeneous 
\cite{bgmr05,bdgm06}.
The main goal of this paper is to formulate a theory of fluctuations for the
PBA model in the dilute limit and to apply it to one of the simplest possible 
state, namely the homogeneous decay state, exploiting its scaling
properties. This will allow us to obtain explicit expressions for the
distributions characterizing the velocity correlations in the system, and
to compute the statistics of the total number of particles, total
momentum and total energy, which 
decrease monotonically due to the annihilation process. 

The paper is organized as follows. In Section \ref{Sec. 2}, we present
the general framework of the hierarchy method \cite{ec81} which allows
to write the evolution equations of correlation functions.  The
specific case of the homogeneous decay state is considered in Section
\ref{Sec. 3}, where the scaling properties of this state are used to
simplify the equations. After
briefly recalling in Section \ref{Sec. 4} how fluctuations and correlations
of global observables can be computed from the knowledge of the
two-particle correlation functions, we first consider
in Section \ref{Sec. 5} the correlation functions at equal time,
which give access to the fluctuations of the total number of
particles, total momentum and total energy. We obtain theoretical
predictions for the asymptotic scaling regime as well as for the short
time behavior, and we test these predictions against numerical
simulations (both Molecular Dynamics and Monte Carlo). 
In Section \ref{Sec. 6}, we generalize our results to the 
two-time correlation functions and compare also to numerical simulations.
Finally, \ref{Sec. 8} contains some discussions of the results and
our conclusions. For the sake of readability, this paper contains
some overlap with its companion \cite{companion}. Repetitions have
been kept to a minimum though, and we therefore refer to \cite{companion}
for several technical details.

\section{General framework}\label{Sec. 2}

The system we consider consists of a dilute gas of identical smooth
hard spheres or disks of mass $m$ and diameter $\sigma$, moving
ballistically in dimension $d$. The particles are supposed to undergo 
only binary collisions. When two particles collide they annihilate with
probability $p$ or collide elastically with probability $1-p$. In this 
probabilistic ballistic annihilation (PBA),
there is therefore no conserved quantity (except for $p=0$) and the number
of particles decreases steadily. 
In this section, we will show how to obtain evolution equations
for the correlation functions in this system. 
The general idea of the method is to derive a closed set of 
equations for the distribution functions describing the fluctuations by using 
the same kind of approximations as needed to derive the kinetic equation, in 
our case the Boltzmann equation. In this way, a unified formalism provides the 
usual kinetic equation as well as evolution equations for the one-time and 
two-time correlations. 

Let $X_j\equiv\{\mathbf{R}_j(t),\mathbf{V}_j(t)\}$ denote the position and
velocity of particle $j$ in the system at time $t$. Both, $\mathbf{R}_j(t)$
and $\mathbf{V}_j(t)$ are parametric functions of the initial positions and
velocities of all particles.
Microscopic one- and two-particle densities in the phase space are defined by
\begin{eqnarray}
F_1(x_1,t)&=&\sum_{i=1}^N\delta[x_1-X_i(t)],\label{i.1}\\
F_2(x_1,x_2,t)&=&\sum_{i=1}^N\sum_{i\ne j}^N \delta[x_1-X_i(t)]
\delta[x_2-X_j(t)], 
\end{eqnarray}
and higher order functions can similarly be defined. Here and
in the following,
the lower case variables $x_i\equiv\{\mathbf{r}_i,\mathbf{v}_i\}$ are 
field variables referring to a particular point in phase space.

The initial state of the system is characterized by a point in phase
space, $\Gamma\equiv\{X_1,\dots,X_N\}$, which is chosen at random
according to a probability $\rho(\Gamma,0)$.  Introducing the notation
$\langle G\rangle\equiv\int\!\!d\Gamma G(\Gamma)\rho(\Gamma,0)$ for
the average over the initial conditions, the
averages of the microscopic densities $F_s(x_1,\dots,x_s,t)$ over  
$\rho(\Gamma,0)$ are the usual one-time reduced distribution functions
\begin{equation}
f_s(x_1,\dots,x_s,t)=\langle F_s(x_1,\dots,x_s,t)\rangle .
\end{equation}
Similarly, 
two-time reduced distribution functions can also be defined in terms of the
microscopic densities as
\begin{equation}
f_{r,s}(x_1,\dots,x_r,t;x_1',\dots,x_s',t')=
\langle F_r(x_1,\dots,x_r,t) F_s(x_1',\dots,x_s',t')\rangle.
\end{equation}
For simplicity we will consider $t>t'>0$ in the following. 

We now introduce the two-particle correlation functions 
through the usual cluster expansion. The one-time 
correlation function $g_2$ and the two-time correlation
function $h_{1,1}$ are then defined by 
\begin{eqnarray}\label{def_g}
 f_2(x_1,x_2,t)=f_1(x_1,t)f_1(x_2,t)+g_2(x_1,x_2,t),\\
\label{def_h}
 f_{1,1}(x_1,t;x_2,t')=f_1(x_1,t)f_1(x_2,t')+h_{1,1}(x_1,t;x_2,t').
\end{eqnarray}
It is easy to show from the definition of $f_1$, $f_2$ and $f_{1,1}$ that
\begin{equation}\label{def.h11}
h_{1,1}(x_1,t;x_2,t)=g_2(x_1,x_2,t)+\delta(x_1-x_2)f_1(x_1,t).
\end{equation}

The case of deterministic annihilation ($p=1$) was considered in
reference \cite{ptd02}. The hierarchy of equations for the reduced 
distribution functions is then shown to be similar to the one describing
elastic collisions, once the binary elastic collision operator
is replaced by the operator describing annihilating collisions.
In the PBA case, assuming molecular chaos, {\it i.e.} that no correlations
exist {\em between colliding particles}, the equation for $f_1(x_1,t)$ 
is  the Boltzmann equation
\begin{equation}\label{ii1}
\left[ \frac{\partial}{\partial t}+L^{(0)}(x_1) \right]f_1(x_1,t)=J[f_1,f_1],
\end{equation}
where 
\begin{eqnarray}
L^{(0)}(x_1)&=&\mathbf{v}_1\cdot\frac{\partial}{\partial\mathbf{r}_1},\\
J[f_1,f_1]&=&\int
dx_2\delta(\mathbf{r}_{12})\bar{T}_0(\mathbf{v}_1,\mathbf{v}_2)
f_1(x_1,t)f_1(x_2,t),
\end{eqnarray}
and
\begin{eqnarray}
\bar{T}_0(\mathbf{v}_1,\mathbf{v}_2)&=&\sigma^{d-1}\int
d\boldsymbol{\hat{\sigma}} \Theta(\mathbf{v}_{12}\cdot
\boldsymbol{\hat{\sigma}})(\mathbf{v}_{12}\cdot
\boldsymbol{\hat{\sigma}})[(1-p)(b_\sigma^{-1}-1)-p],
\end{eqnarray}
is the PBA binary collision operator. In the above expressions 
$\mathbf{v}_{12}=\mathbf{v}_{1}-\mathbf{v}_{2}$ is the relative velocity,
$\mathbf{r}_{12}=\mathbf{r}_{1}-\mathbf{r}_{2}$ the relative position,
$\Theta$ is the Heaviside function,  $\boldsymbol{\hat{\sigma}}$ a unit
vector joining the centers of two particles at collision, and
$b_\sigma^{-1}$ is an operator that replaces all the velocities
$\mathbf{v}_1$ and $\mathbf{v}_2$ appearing to its right by the
precollisional values $\mathbf{v}_1^*$ and $\mathbf{v}_2^*$
\begin{eqnarray}\label{op_bmenos1v1}
b_\sigma^{-1}\mathbf{v}_1\equiv\mathbf{v}_1^*&=&\mathbf{v}_1
-(\mathbf{v}_{12}\cdot\boldsymbol{\hat{\sigma}})\boldsymbol{\hat{\sigma}},\\
\label{op_bmenos1v2}
b_\sigma^{-1}\mathbf{v}_2\equiv\mathbf{v}_2^*&=&\mathbf{v}_2
+(\mathbf{v}_{12}\cdot\boldsymbol{\hat{\sigma}})\boldsymbol{\hat{\sigma}}.
\end{eqnarray}
The equation for the correlation function $g_2$ can be obtained 
under the same hypothesis required to derive the Boltzmann equation,
following the same lines as in reference \cite{bgmr04} in the case of
inelastically colliding particles, as
\begin{eqnarray}\label{ii.6}
\left[\frac{\partial}{\partial t}+L^{(0)}(x_1)+L^{(0)}(x_2)-K(x_1,t)-K(x_2,t)\right]
g_2(x_1,x_2,t)\nonumber\\
=\delta(\mathbf{r}_{12})\bar{T}_0(\mathbf{v}_1,\mathbf{v}_2)
f_1(x_1,t)f_1(x_2,t) ,
\end{eqnarray}
where we have introduced the linear operator $K(x_i,t)$ 
\begin{equation}\label{ii.5}
K(x_i,t)=\int dx_3\delta(\mathbf{r}_{i3})\bar{T}_0(\mathbf{v}_i,\mathbf{v}_3)
(1+{\cal P}_{i3})f_1(x_3,t),
\end{equation}
and where the permutation operator  ${\cal P}_{ab}$ interchanges the labels of
particles $a$ and $b$ in the quantities on which it acts. 

Finally, the evolution equation for $h_{1,1}$ reads
\begin{equation}
\left[\frac{\partial}{\partial t}+L^{(0)}(x_1)-K(x_1,t)\right]
h_{1,1}(x_1,t;x_2,t')=0, \qquad t>t', 
\end{equation}
that has to be solved with the initial condition (\ref{def.h11}),
$h_{1,1}(x_1,t';x_2,t')=g_2(x_1,x_2,t')+\delta(x_1-x_2)f_1(x_1,t')$.

The equations for the correlation functions $h_{1,1}$ and $g_2$ contain a part
corresponding to free streaming and another one which corresponds to
collisions. In particular, the evolution of the one-time correlation function $g_2$
due to collisions has two parts: one due to collisions of particle $1$ or $2$
(corresponding to the indices of the correlation function) with a third
particle, which is governed by the Boltzmann collision operator; and a second
one, due to collisions of particle $1$ with particle $2$, which can be written
in terms of the one particle distribution function as a consequence of the
molecular chaos hypothesis.  In fact, as in the case of the inelastic
granular gas \cite{bgmr04}, the only difference between the evolution
equations of the correlation functions for the PBA and for a system of elastic 
particles lies in the substitution of the elastic binary
collision operator by the operator for the PBA model,
$\bar{T}_0(\mathbf{v}_1,\mathbf{v}_2)$. However this does not give any a
priori guarantee on the range of validity of these equations. This
prescription, {\it i.e.} how small the density of the system must be so that
the above kinetic 
equations provide an accurate description, might depend on the parameter $p$,
and also on the particular state being considered.

We will see in Section \ref{Sec. 4} how the knowledge of $h_{1,1}$ and
$g_2$ allows to obtain the correlation functions of any observable
which is a function of the particles positions in phase space.

\section{Homogeneous Decay State}\label{Sec. 3}

As recalled in the previous companion paper \cite{companion}, the Boltzmann
equation for the PBA model (\ref{ii1}) admits a particular solution
$f_{H}(\mathbf{v},t)$ describing a spatially homogeneous decay state
(HDS), in which all the time dependence is contained in the evolution
of the density $n_H(t)$ and the temperature $T_H(t)$, which are defined as
in standard Kinetic Theory as
\begin{equation}
n_H(t)=\int\!\!d\mathbf{v}f_H(\mathbf{v},t),\qquad 
\frac{d}{2}n_H(t)T_H(t)=\int\!\!d\mathbf{v}\frac{m}{2}v^2f_H(\mathbf{v},t).
\end{equation}
Although there exists no
rigorous proof of its stability nor of the fact that such a state
should be approached from arbitrary initial conditions, numerical 
results obtained by 
Molecular Dynamic simulations and by the direct
Monte Carlo method strongly support the existence of such a 
homogeneous solution \cite{cdt04s,t02}. In this section,
we review for completeness the evolution equation of the 
one-particle distribution function and obtain the equations for
adequately rescaled correlation functions. All quantities concerning
this homogeneous decay state will be labeled by an index $H$.

In the HDS, the one body distribution function does not depend
on space and follows the scaling form \cite{ptd02}
\begin{equation}\label{iii.1}
f_{H}(\mathbf{v},t)=\frac{n_H(t)}{v_H^{d}(t)}\chi_H(\mathbf{c}),
\end{equation}
where $n_H(t)$ is the uniform density, $v_H(t)$ is the thermal
(root-mean-square) velocity related to the granular temperature 
$T_H(t)$ by
\begin{equation}
v_H(t)=\left[\frac{2T_H(t)}{m}\right]^{1/2},
\end{equation}
and $\chi(\mathbf{c})$ is an isotropic function depending only on the
modulus $c=|\mathbf{c}|$ of the rescaled velocity
$\mathbf{c}=\mathbf{v}/v_H(t)$. Moreover, the 
density and temperature fields evolve according to \cite{cdt04}
\begin{eqnarray}\label{evolucion_N}
\frac{dn_H(t)}{dt}&=&-p\nu_H(t)\zeta _n n_H(t),\\
\label{evolucion_E}
\frac{dT_H(t)}{dt}&=&-p\nu_H(t)\zeta _T T_H(t),
\end{eqnarray}
where $\nu_H(t)$ is the collision frequency 
of the corresponding hard sphere fluid in equilibrium
(with same temperature and density)
\begin{equation}
\nu_H(t)=\frac{n_H(t) \,T^{1/2}_H(t) \,\sigma^{d-1}}{ m^{1/2}}
\frac{8\pi^{\frac{d-1}{2}}}{(d+2)\Gamma(d/2)} , 
\end{equation}
and the dimensionless decay rates $\zeta_n$ and $\zeta_T$ 
are functionals of the distribution function:
\begin{eqnarray}\label{zeta_n}
p\zeta_n&=&-\frac{\gamma}{2}\int\!\!d\mathbf{c}_1\!\!\int\!\!d\mathbf{c}_2
T(\mathbf{c}_1,\mathbf{c}_2)\chi_H(\mathbf{c}_1)\chi_H(\mathbf{c}_2),\\
\label{zeta_T}
p\zeta_T&=&-\frac{\gamma}{2}\int\!\!d\mathbf{c}_1\!\!\int\!\!d\mathbf{c}_2
\left(\frac{2c_1^2}{d}-1\right)T(\mathbf{c}_1,\mathbf{c}_2)
\chi_H(\mathbf{c}_1)\chi_H(\mathbf{c}_2).
\end{eqnarray}
In these expressions, $\gamma$, which does not depend on time,
is given by
$\gamma=2n_H(t)v_H(t)\sigma^{d-1}/\nu_H(t)=
(d+2)\sqrt{2}\Gamma(d/2)/\left(4\pi^{(d-1)/2}\right)$,
and the binary collision operator 
$T(\mathbf{c}_1,\mathbf{c}_2)$ takes the form
\begin{equation}
T(\mathbf{c}_1,\mathbf{c}_2)=\int\!\!d\sigmagorro
\Theta(\mathbf{c}_{12}\cdot\sigmagorro)(\mathbf{c}_{12}\cdot\sigmagorro)
[(1-p)b_{\sigma}^{-1}-1]. 
\end{equation}
Finally, the scaled distribution function 
$\chi_H(\mathbf{c})$ obeys the equation
\begin{equation}\label{ec_chi}
p\left[(d\zeta_T-2\zeta_n)+\zeta_T\mathbf{c}_1\cdot
\frac{\partial}{\partial\mathbf{c}_1}\right]\chi_H(\mathbf{c}_1)=\gamma\int\!\!
d\mathbf{c}_2T(\mathbf{c}_1,\mathbf{c}_2)\chi_H(\mathbf{c}_1)
\chi_H(\mathbf{c}_2). 
\end{equation}
The operator $b_\sigma^{-1}$ in the last equation is defined again by equations
(\ref{op_bmenos1v1}) and (\ref{op_bmenos1v2}), but substituting 
$(\mathbf{v}_1,\mathbf{v}_2)$ by $(\mathbf{c}_1,\mathbf{c}_2)$. 
The analytical form of $\chi_H$ is not known, but its behavior at 
large and small
velocities has been determined \cite{t02,ptd02}. As in the companion
paper \cite{companion}, we will use here the approximate form of the
distribution function in the so-called first Sonine approximation, which is
valid for velocities in the thermal region, and all the functionals
of $\chi_H(\mathbf{c})$, like the decay rates and the transport 
coefficients, will be evaluated in this approximation \cite{cdt04s,t02}.

Considering the scaling form for the one-particle distribution function, 
it is convenient to introduce the rescaled correlation function
$\tilde{g}_H$ through
\begin{equation}\label{iii.11}
g_{2,H}(x_1,x_2,t)=\frac{n_H(t)}{v_H^{2d}(t)}
\tilde{g}_H(\tau,\mathbf{r}_{12},\mathbf{c}_1,\mathbf{c}_2),
\end{equation}
where we have taken into account that the system is invariant under space
translation, so that $g_{2,H}$ depends on
$\mathbf{r}_{12}=\mathbf{r}_{1}-\mathbf{r}_{2}$ and not on 
$\mathbf{r}_{1}$ and $\mathbf{r}_{2}$ separately.
The dimensionless time scale $\tau$
\begin{equation}
\tau=\frac{1}{2}\int_0^tdt'\nu_H(t'),
\end{equation}
is proportional to the number of collisions in the time interval $[0,t]$.
The equation for the reduced function $\tilde{g}_H$ in these
units reads then
\begin{eqnarray}\label{iii.12}
\left[-\frac{\partial}{\partial\tau}
+\Lambda(\mathbf{c}_1)+\Lambda(\mathbf{c}_2)
-2p\zeta_n -l_H(\tau)\mathbf{c}_{12}\cdot\frac{\partial}
{\partial\mathbf{r}_{12}}
\right]\tilde{g}_H(\tau,\mathbf{r}_{12},\mathbf{c}_1,\mathbf{c}_2)\nonumber\\
=-\delta(\mathbf{r}_{12})
\gamma T(\mathbf{c}_1,\mathbf{c}_2)\chi_H(\mathbf{c}_1)\chi_H(\mathbf{c}_2),
\end{eqnarray}
where we have also introduced the length scale
$l_H(t)=2v_H(t)/\nu_H(t)$, which is proportional to the
instantaneous mean free path,
and the linearized Boltzmann operator (see previous paper)
\begin{eqnarray}
\Lambda(\mathbf{c}_i)h(\mathbf{c}_i)&=&\gamma \int\!\!d\mathbf{c}_3
T(\mathbf{c}_i,\mathbf{c}_3)(1+{\cal P}_{i3})
\chi_H(\mathbf{c}_3)h(\mathbf{c}_i)
\nonumber\\
&+&p(2\zeta_n -d\zeta_T )h(\mathbf{c}_i)-p\zeta_T \mathbf{c}_i\cdot
\frac{\partial}{\partial\mathbf{c}_i}h(\mathbf{c}_i).
\end{eqnarray}

Similarly, we define a rescaled two-time correlation function 
$\tilde{h}_H$  through
\begin{equation}
h_{1,1,H}(x_1,t;x_2,t')=\frac{n_H(t)}
{v_H^d(t)v_H^d(t')}\tilde{h}_H
(\mathbf{r}_{12};\mathbf{c}_1,\tau;\mathbf{c}_2,\tau') , 
\end{equation}
and obtain the following evolution equation: 
\begin{equation}\label{ec.ev.h11}
\frac{\partial}{\partial\tau}\tilde{h}_H(\mathbf{r}_{12};\mathbf{c}_1,\tau;
\mathbf{c}_2,\tau')=\left[\Lambda(\mathbf{c}_1)-
l_H(\tau) \mathbf{c}_1\cdot\frac{\partial}{\partial\mathbf{r}_1}\right]
\tilde{h}_H(\mathbf{r}_{12};\mathbf{c}_1,\tau;\mathbf{c}_2,
\tau')  ,
\end{equation}
with $\tilde{h}_H(\mathbf{r}_{12};\mathbf{c}_1,\tau;\mathbf{c}_2,\tau)=
\tilde{g}_H(\tau,\mathbf{r}_{12},\mathbf{c}_1,\mathbf{c}_2)
+ \delta(\mathbf{c}_1 - \mathbf{c}_2)
\delta(\mathbf{r}_{12})\chi_H(\mathbf{c}_1)$.

It is interesting to note how,
in this representation, all the time dependence due to the reference state is 
absorbed in the free streaming term through the function $l_H(\tau)$,
proportional to the mean free path. The evolution of the correlation
functions moreover will be determined by the properties of the
linearized Boltzmann operator $\Lambda$, which we have already studied
in the companion paper \cite{companion} and which we will recall in Section \ref{Sec. 5}.

\section{From particle correlation functions to 
correlations and
fluctuations of global magnitudes}
\label{Sec. 4}

In this section, we will show how the previously presented framework
for correlation functions will allow us to study the fluctuations and
correlations of global quantities for a PBA system in the
homogeneous decay state.  In
particular, we will focus on the total number of particles $N$, the
total momentum $\mathbf{P}$, and the total energy $E$.

Consider indeed two dynamical variables of the form
\begin{eqnarray}\nonumber
A\left[\Gamma(t)\right]=\sum_{i=1}^Na(\mathbf{V}_i)=
\int\!\!dx_1a(\mathbf{v}_1)F_1(x_1,t) \\
\label{iv.4}\label{abF}
B\left[\Gamma(t)\right]=\sum_{i=1}^Nb(\mathbf{V}_i)=
\int\!\!dx_2b(\mathbf{v}_2)F_1(x_2,t)
\end{eqnarray}
where $a$ and $b$ are functions of the particles' velocities
$\mathbf{V}_i$, and $F_1(x_1,t)$ is the microscopic density in
phase space (\ref{i.1}). Taking $a=1$, $a=\mathbf{v}$ and $a=m
\mathbf{v}_i^2/2$ yield for $A$ the total number of particles $N$, the
total momentum $\mathbf{P}$, and the total kinetic energy $E$, respectively. The
deviations $\delta A(t)=A(t)-\langle A(t)\rangle_H$ and $\delta
B(t)=B(t)-\langle B(t)\rangle_H$ of $A$ or $B$ from their average
values in the HDS (denoted by $\langle\dots\rangle_H$), define the
fluctuations of both magnitudes, which have average zero and
correlations
\begin{equation}\label{c.5}
\langle \delta A(t)\delta B(t')\rangle_H=\langle
A(t)B(t')\rangle_H-\langle A(t)\rangle_H\langle B(t')\rangle_H.
\end{equation} 
It is then straightforward to use the definition of 
the two-time correlation function $h_{1, 1}$ in Eq. (\ref{def_h}), to obtain
\begin{equation} \label{eq2times}
\langle\delta
A(t)\delta B(t^\prime)\rangle_H=\int\!\!d\mathbf{r}_1\!\!\int\!\!
d\mathbf{v}_1\!\!\int\!\!d\mathbf{r}_2\!\!\int\!\!d\mathbf{v}_2
a(\mathbf{v}_1)b(\mathbf{v}_2)
h_{1,1,H}(\mathbf{r}_1,\mathbf{v}_1,t;\mathbf{r}_2,\mathbf{v}_2,t^\prime).
\end{equation}
In particular, for $t=t'$, this leads to
\begin{eqnarray}\label{iv.5}
\langle \delta A(t)\delta B(t)\rangle_H&=&V\int\!\!d\mathbf{v} a(\mathbf{v})
b(\mathbf{v}) f_H(\mathbf{v},t)\nonumber\\
&+&V\int\!\!d\mathbf{v}_1\!\!\int\!\!d\mathbf{v}_2
a(\mathbf{v}_1)b(\mathbf{v}_2)\int\!\!d\mathbf{r}_{12}
g_{2,H}(\mathbf{r}_{12},\mathbf{v}_1,\mathbf{v}_2,t) ,
\end{eqnarray}
where $V=\int\!d\mathbf{r}_1$ is the total volume of the system.
These formulas show how the correlations of two different global
magnitudes are determined by the one particle 
distribution function and by the correlation functions. The one particle 
distribution function is known in the HDS in the first Sonine approximation 
\cite{cdt04s,t02}.

\section{Fluctuations in the HDS}
\label{Sec. 5}

Let us focus in this section on the one-time correlation function
$\tilde{g}_H$. We will only need functions $a$ and $b$ which depend
on the velocity degrees of freedom, so that
it is convenient to integrate out the spatial dependence
by introducing
\begin{equation}
\phi_{H}(\tau,\mathbf{c}_1,\mathbf{c}_2)\equiv \int\!\!d\mathbf{r}_{12}
\tilde{g}_H(\tau,\mathbf{r}_{12},\mathbf{c}_1,\mathbf{c}_2) ,
\end{equation}
whose evolution is obtained from (\ref{iii.12}) as
\begin{equation}\label{eq.phi}
\left[-\frac{\partial}{\partial\tau}+\Lambda(\mathbf{c}_1)
+\Lambda(\mathbf{c}_2)-2p\zeta_n \right]
\phi_{H}(\tau,\mathbf{c}_1,\mathbf{c}_2)
=-\gamma T(\mathbf{c}_1,\mathbf{c}_2)\chi_H(\mathbf{c}_1)\chi_H(\mathbf{c}_2).
\end{equation}
Given an initial condition $\phi_H(0,\mathbf{c}_1,\mathbf{c}_2)$, this equation
(\ref{eq.phi}) can be formally integrated as
\begin{eqnarray}\label{formal}
\phi_H(\tau,\mathbf{c}_1,\mathbf{c}_2)&=&e^{(\Lambda(\mathbf{c}_1)+\Lambda(\mathbf{c}_2)-2p\zeta_n)
\tau}\phi_{H}(0,\mathbf{c}_1,\mathbf{c}_2)\nonumber\\
&+&\int_0^\tau d\tau^\prime e^{(\tau-\tau^\prime)
[\Lambda(\mathbf{c}_1)+\Lambda(\mathbf{c}_2)-2p\zeta_n]}\gamma
T(\mathbf{c}_1,\mathbf{c}_2)\chi_H(\mathbf{c}_1)\chi_H(\mathbf{c}_2)\nonumber\\
&=&e^{(\Lambda(\mathbf{c}_1)+\Lambda(\mathbf{c}_2)-2p\zeta_n)\tau}
\left[\phi_H(0,\mathbf{c}_1,\mathbf{c}_2)-\phi^s_H(\mathbf{c}_1,\mathbf{c}_2)
\right]+{\phi}^s_H(\mathbf{c}_1,\mathbf{c}_2).\nonumber\\
\end{eqnarray}
where ${\phi}^s_H(\mathbf{c}_1,\mathbf{c}_2)$ is the solution of
\begin{equation}\label{iv.2}
\left[\Lambda(\mathbf{c}_1)+\Lambda(\mathbf{c}_2)-2p\zeta_n \right]
\phi^s_{H}(\mathbf{c}_1,\mathbf{c}_2)
=-\gamma T(\mathbf{c}_1,\mathbf{c}_2)\chi_H(\mathbf{c}_1)\chi_H(\mathbf{c}_2),
\end{equation}
where we implicitly assumed that $\Lambda$ is invertible.
This happens to be the case, see below.
The spectral properties of the linearized Boltzmann
operator $\Lambda$ are thus crucial for the evaluation of
$\phi_{H}$. We therefore start by recalling these properties.

\subsection{Spectral properties of $\Lambda$}

In our companion paper \cite{companion}, we have analyzed the eigenvalue problem associated
with the linearized Boltzmann operator~$\Lambda$
\begin{equation}
\Lambda(\mathbf{c})\xi_\beta(\mathbf{c})=\lambda_\beta\xi_\beta(\mathbf{c}).
\end{equation}
We have in fact restricted ourselves to the hydrodynamic part of $\Lambda$,
defined, by those
eigenvalues of the balance equations for the number density, momentum, and
temperature following from the homogeneous linearized Boltzmann
equation. Such eigenvalues are \cite{cdt04}
\begin{equation}
\lambda_1=0,\qquad\lambda_2=-p(\zeta_T +2\zeta_n ),\qquad\lambda_3=
p\zeta_T .
\end{equation}
Although we were not able to prove that these eigenvalues are
indeed the hydrodynamic ones, the 
self-consistency of the resulting description and the successful comparison
with numerical simulations have validated this assumption \cite{companion}.
In the previous paper we also obtained the corresponding 
eigenfunctions 
\begin{eqnarray}\label{eigenfunction1}
\xi_1(\mathbf{c})&=&\chi_H(\mathbf{c}_1)
+\frac{\partial}{\partial \mathbf{c}}\cdot[\mathbf{c}\chi_H(\mathbf{c})],\\
\xi_2(\mathbf{c})&=&z\chi_H(\mathbf{c})
-\frac{\partial}{\partial\mathbf{c}}\cdot[\mathbf{c}\chi_H(\mathbf{c})],\\
\label{eigenfunction3}
\boldsymbol{\xi}_3(\mathbf{c})&=&
-\frac{\partial\chi_H(\mathbf{c})}{\partial\mathbf{c}},
\end{eqnarray}
with $z=\frac{2\zeta_n }{\zeta_T }$ a function of the probability of
annihilation $p$. The eigenvalue $\lambda_3$ is $d$-fold degenerated and we denote $\xi_{3i}$, $i=1,...,d$ the corresponding eigenvectors.
The scalar
product of two functions $f(\mathbf{c})$ and $g(\mathbf{c})$ is defined as
\begin{equation}\label{escalar-product1}
\langle f\mid g\rangle\equiv\int\!\!d\mathbf{c}\chi_H^{-1}(\mathbf{c})
f^*(\mathbf{c})g(\mathbf{c}),
\end{equation}
$f^*$ being the complex conjugate of $f$. The eigenfunctions $\xi_\beta$
given in (\ref{eigenfunction1})-(\ref{eigenfunction3}) are not orthogonal, 
as a consequence of the
operator $\Lambda$ being non-Hermitian in the associated Hilbert space. 
On the other hand, it is easily verified that the set of functions
$\{\bar{\xi}_1;\bar{\xi}_2;\boldsymbol{\bar{\xi}}_3\}=
\left\{\chi_H(\mathbf{c})-\frac{z}{1+z}\left(\frac{1}{2}
+\frac{c^2}{d}\right)\chi_H(\mathbf{c});\frac{1}{1+z}\left(\frac{1}{2}
+\frac{c^2}{d}\right)\chi_H(\mathbf{c});\mathbf{c}\chi_H(\mathbf{c})\right\}$
verify the biorthogonality condition $
\langle\bar{\xi}_\beta\vert\xi_{\beta'}\rangle=\delta_{\beta, \beta'}$,
for $\beta$, $\beta'=1,2,3$.

The eigenfunctions of the operator
$\left[\Lambda(\mathbf{c}_1)+\Lambda(\mathbf{c}_2)-2 p\zeta_n \right]$ that 
appear in equation (\ref{iv.2}) are then easily seen to be
$\xi_{\beta_1}(\mathbf{c}_1)\xi_{\beta_2}(\mathbf{c}_2)$, with
\begin{equation}
\left[\Lambda(\mathbf{c}_1)+\Lambda(\mathbf{c}_2)-2 p\zeta_n \right]
\xi_{\beta_1}(\mathbf{c}_1)\xi_{\beta_2}(\mathbf{c}_2)
=(\lambda_{\beta_1}+\lambda_{\beta_2}-2p\zeta_n )
\xi_{\beta_1}(\mathbf{c}_1)\xi_{\beta_2}(\mathbf{c}_2) .
\end{equation}
Since $\zeta_n >\zeta_T $ \cite{cdt04}, and under the assumption that the norm
of the ``non hydrodynamic'' eigenvalues are always greater than the
hydrodynamic ones, the eigenvalues of
$\left[\Lambda(\mathbf{c}_1)+\Lambda(\mathbf{c}_2)-2 p\zeta_n \right]$ 
are therefore all negatives. This has the important consequence
that the exponential term in (\ref{formal}) decays to zero
and that the large time limit of $\phi_H(\tau,\mathbf{c}_1,\mathbf{c}_2)$ is
 ${\phi}^s_H(\mathbf{c}_1,\mathbf{c}_2)$, solution of Eq. (\ref{iv.2}).

\subsection{Hydrodynamic part of the correlation functions}

Obtaining the full spectrum of $\Lambda$ is a formidable task.
We will here assume, as in the companion paper \cite{companion}, that the 
kinetic (non-hydrodynamic) modes have a fast decay, and work
in the hydrodynamic subspace spanned by the functions 
$\xi_\beta$ defined in the previous subsection. In that purpose, we
generalize 
the definition of the scalar product given in (\ref{escalar-product1})
to two-velocity functions by
\begin{equation}
\langle f(\mathbf{c}_1,\mathbf{c}_2)\vert
g(\mathbf{c}_1,\mathbf{c}_2)\rangle\equiv\int\!\!d\mathbf{c}_1\!\!\int\!\!
d\mathbf{c}_2 \chi_H^{-1}(\mathbf{c}_1)\chi_H^{-1}(\mathbf{c}_2)
f^*(\mathbf{c}_1,\mathbf{c}_2)g(\mathbf{c}_1,\mathbf{c}_2).
\end{equation}
This allows to define a projector operator $P_{12}$ onto the space
spanned by the functions 
$\xi_{\beta_1}(\mathbf{c}_1)\xi_{\beta_2}(\mathbf{c}_2)$ as
\begin{equation}
P_{12}f(\mathbf{c}_1,\mathbf{c}_2)\equiv\sum_{\beta_1=1}^3\sum_{\beta_2=1}^3
\langle \bar{\xi}_{\beta_1}(\mathbf{c}_1)\bar{\xi}_{\beta_2}(\mathbf{c}_2)
\vert f(\mathbf{c}_1,\mathbf{c}_2)\rangle 
\xi_{\beta_1}(\mathbf{c}_1)\xi_{\beta_2}(\mathbf{c}_2),
\end{equation}
and the ``hydrodynamic part'' of $\phi_H(\tau,\mathbf{c}_1,\mathbf{c}_2)$ 
and ${\phi}^s_H(\mathbf{c}_1,\mathbf{c}_2)$
are by definition 
\begin{eqnarray}
\phi_H^{(h)}(\tau,\mathbf{c}_1,\mathbf{c}_2)\equiv P_{12}
\phi_H(\tau,\mathbf{c}_1,\mathbf{c}_2) =
\sum_{\beta_1,\beta_2=1}^3
a_{\beta_1,\beta_2}(\tau)
\xi_{\beta_1}(\mathbf{c}_1)\xi_{\beta_2}(\mathbf{c}_2),\\
\phi_H^{s\,(h)}(\mathbf{c}_1,\mathbf{c}_2)\label{b.15}
\equiv P_{12}{\phi}^s_H(\mathbf{c}_1,\mathbf{c}_2)=\sum_{\beta_1,\beta_2=1}^3
a^s_{\beta_1,\beta_2}
\xi_{\beta_1}(\mathbf{c}_1)\xi_{\beta_2}(\mathbf{c}_2) .
\end{eqnarray}

We can now obtain a closed equation for $\phi_H^{(h)}$ by 
applying the operator $P_{12}$ on both sides of equations 
(\ref{formal}) and (\ref{iv.2}), under the additional assumption
that
\begin{equation}\label{aproximacionPL}
P_{12}\Lambda(\mathbf{c}_i)=P_{12}\Lambda(\mathbf{c}_i)P_{12}.
\end{equation}
A theoretical estimation ``a priori'' of the accuracy of this approximation
would require to know more than it is available at present
about the spectrum of $\Lambda(\mathbf{c})$ and its
adjoint. Therefore, it will be taken as a
working hypothesis to be validated later on by comparing the predictions it
leads to with the results from numerical simulations of the system. 
It is worth emphasizing that since $\Lambda$ leaves the hydrodynamic subspace
invariant, Equation (\ref{aproximacionPL}) is equivalent to the commutation relation
$P_{12}\Lambda = \Lambda P_{12}$.
Proceeding further, we obtain from (\ref{formal})
\begin{eqnarray}\label{ec.Pphi_tau}
\phi_H^{(h)}(\tau,\mathbf{c}_1,\mathbf{c}_2)
&=&e^{(\Lambda(c_1)+\Lambda(c_2)-2p\zeta_n)\tau}P_{12}
\left[\phi_H(0,\mathbf{c}_1,\mathbf{c}_2)-{\phi}^s_H(\mathbf{c}_1,\mathbf{c}_2)
\right]+P_{12}{\phi}^s_H(\mathbf{c}_1,\mathbf{c}_2)\nonumber\\
&=&\sum_{\beta_1,\beta_2=1}^3\left[A_{\beta_1,\beta_2}
e^{(\lambda_{\beta_1}+\lambda_{\beta_2}-2p\zeta_n)\tau}+a^s_{\beta_1,\beta_2}
\right] \xi_{\beta_1}(\mathbf{c}_1)\xi_{\beta_2}(\mathbf{c}_2),
\end{eqnarray}
where we have introduced
$A_{\beta_1,\beta_2}=a_{\beta_1,\beta_2}(0)-a^s_{\beta_1,\beta_2}$.
We show in Appendix \ref{appendixA} 
how to obtain explicit formulas for the coefficients 
$a^s_{\beta_1,\beta_2}$ in terms of the cooling rates, $\zeta_n $ and
$\zeta_T $, and other coefficients which are also functionals of the one time
distribution function $\chi_H$. 
The values of $a_{\beta_1,\beta_2}(0)$ depend on the initial
  condition $\phi_H(0,\mathbf{c}_1,\mathbf{c}_2)$. For the specific
  case in which the 
  variables $N$,  
$\mathbf{P}$ and $E$ do not fluctuate at $t=0$, and taking into
  account that the system is in the HDS, the coefficients  
$a_{\beta_1,\beta_2}(0)$ are calculated in Appendix \ref{appendixB}. 

\subsection{Hydrodynamic approximation for global fluctuations}

 In this section we compute the correlation functions of
  the global observables by replacing $\phi_H$ implicitly appearing in 
(\ref{iv.5}) by its hydrodynamic part, $\phi_H^{(h)}$. This can be
  done invoking the relation
\begin{equation}
\langle\bar{\xi}_{\beta_1}(\mathbf{c}_1)\bar{\xi}_{\beta_2}(\mathbf{c}_2)
\vert f(\mathbf{c}_1,\mathbf{c}_2)\rangle=
\langle\bar{\xi}_{\beta_1}(\mathbf{c}_1)\bar{\xi}_{\beta_2}(\mathbf{c}_2)
\vert f^{(h)}(\mathbf{c}_1,\mathbf{c}_2)\rangle, 
\end{equation}
for $\beta_i=1,2,3$. However, it must be stressed that the theoretical
prediction for $\phi_H^{(h)}$ in Eq. (\ref{ec.Pphi_tau}), has been
calculated using the approximation (\ref{aproximacionPL}).

If we substitute $a(\mathbf{v})=1$ and $b(\mathbf{v})=1$ in (\ref{iv.5}), we 
obtain for the fluctuations of the number of particles
\begin{equation}\label{c.7}
\langle\delta N^2(\tau)\rangle_H=N_H(\tau)\left[\int\!\!
d\mathbf{c}\chi_H(\mathbf{c})+\int\!\!d\mathbf{c}_1\!\!\int\!\!
d\mathbf{c}_2 \phi_{H}^{(h)}(\tau,\mathbf{c}_1,\mathbf{c}_2)\right],
\end{equation}
where  we have introduced the notation $N_H\equiv \langle N\rangle_H$.
In order to calculate the fluctuations of the total momentum we substitute
$a(\mathbf{v})=v_i$ and $b(\mathbf{v})=v_j$ in equation (\ref{iv.5}) and
obtain
\begin{eqnarray}
\langle\delta P_i(\tau)\delta P_j(\tau)\rangle_H=
%\qquad\qquad\qquad\qquad\qquad\qquad\qquad\qquad\qquad\qquad\qquad\qquad
%\nonumber\\
%\qquad\qquad
N_H(\tau)v_H^2(\tau)\left[\int\!\!d\mathbf{c} c_ic_j\chi_H(\mathbf{c})
+\int\!\!d\mathbf{c}_1\!\!\int\!\!d\mathbf{c}_2
c_{1i}c_{2j}\phi_{H}^{(h)}(\tau,\mathbf{c}_1,\mathbf{c}_2)\right].
\end{eqnarray}
For the energy we substitute $a(\mathbf{v})=b(\mathbf{v})=\frac{1}{2}mv^2$
so that we have
\begin{equation}
\langle\delta E^2(\tau)\rangle_H=\frac{m^2}{4}N_H(\tau)v_H^4(\tau)
\left[\int\!\!d\mathbf{c}
c^4\chi_H(\mathbf{c})+\int\!\!d\mathbf{c}_1\!\!\int\!\!d\mathbf{c}_2
c_1^2c_2^2\phi_{H}^{(h)}(\tau,\mathbf{c}_1,\mathbf{c}_2)\right].
\end{equation}
Finally, we can calculate the correlation between $\delta N$ and $\delta E$
by taking $a(\mathbf{v})=1$ and $b(\mathbf{v})=\frac{1}{2}mv^2$
\begin{equation}
\langle\delta N(\tau)\delta E(\tau)\rangle_H
=\frac{m}{2}N_H(\tau)v_H^2(\tau)\left[\int\!\!d\mathbf{c}
c^2\chi_H(\mathbf{c})+\int\!\!d\mathbf{c}_1\!\!\int\!\!d\mathbf{c}_2
c_2^2\phi_{H}^{(h)}(\tau,\mathbf{c}_1,\mathbf{c}_2) \right].
\end{equation}

After some algebra, we obtain
\begin{eqnarray}
&&\langle\delta N^2(\tau)\rangle_H=N_H(\tau)
\left[1+a^s_{1,1}+2za^s_{1,2}+z^2 a^s_{2,2}+ A_{1,1} e^{-2p\zeta_n\tau} 
%\right.
%\nonumber\\
%\qquad\qquad\qquad
+
%\left.
2zA_{1,2}e^{-p(\zeta_T+4\zeta_n)\tau}+z^2A_{2,2}
e^{-2p(\zeta_T+3\zeta_n)\tau}\right], \label{prediction1} \\
&&\langle\delta
P_i(\tau) \delta
P_j(\tau)\rangle_H=\delta_{ij} 
N_H(\tau)v_H^2(\tau)\left(\frac{1}{2}+a^s_{3i,3i}\right) 
\left[1-e^{-2p(\zeta_n-\zeta_T)\tau}\right],
\end{eqnarray}
\begin{eqnarray}
\langle\delta E^2(\tau)\rangle_H=\frac{m^2}{4}N_H(\tau)v_H^4(\tau)
&&\left[\frac{d(d+2)}{4}(1+a_2)+\frac{d^2}{4}a^s_{1,1}
\right.
%\qquad\qquad\nonumber\\
-d^2(1+\frac{z}{2})a^s_{1,2}+d^2(1+\frac{z}{2})^2
+\frac{d^2}{4}A_{1,1}e^{-2p\zeta_n\tau}\nonumber\\
%\qquad\qquad
&&-d^2\left(1+\frac{z}{2}\right)
\left.
A_{1,2}e^{-p(\zeta_T+4\zeta_n)\tau}+d^2
\left(1+\frac{z}{2}\right)^2A_{2,2}e^{-2p(\zeta_T+3\zeta_n)\tau}\right],
\label{prediction2} 
\end{eqnarray}
and
\begin{eqnarray}
\langle\delta N(\tau)\delta E(\tau)\rangle_H=\frac{m}{2}N_H(\tau)v_H^2(\tau)
&&\left[\frac{d}{2}-\frac{d}{2}a^s_{1,1}+da^s_{1,2}
+dz \left(1+\frac{z}{2}\right)
a^s_{2,2}\right. -\frac{d}{2}A_{1,1}e^{-2p\zeta_n\tau}\nonumber\\ 
&&+\left.dA_{1,2}e^{-p(\zeta_T+4\zeta_n)\tau}+
d z \left(1+\frac{z}{2} \right)A_{2,2}
e^{-2p(\zeta_T+3\zeta_n)\tau}\right],\label{prediction3}
\end{eqnarray}
where $a_2$ is related to the fourth moment of $\chi_H(\mathbf{c})$ 
through $\int d\mathbf{c} c^4\chi_H(\mathbf{c})=\frac{d(d+2)}{4}[1+a_2]$,
and has been evaluated
in the first Sonine approximation in \cite{ptd02}.
All the functions $a^s_{\alpha,\beta}$ and $A_{\alpha,\beta}$ are evaluated in
the Appendices \ref{appendixA} and \ref{appendixB}.

At this point, it is important to note that the equations
(\ref{prediction1})-(\ref{prediction3}) have been obtained under the
assumption that the system is in the homogeneous decay state at all times, {\it i.e.} that
the one particle distribution function is $\chi_H(\mathbf{c})$ for all the
time evolution. For $p<1$, if we start with an arbitrary initial 
condition, numerical simulations show that, after a few collisions, the 
distribution function reaches 
the scaling regime given by equation (\ref{iii.1}). Then, the evolution of 
$\phi_H^{(h)}$ is given by (\ref{ec.Pphi_tau}) and one expects that
the same correlation functions (\ref{prediction1})-(\ref{prediction3})
will be obtained in the long time 
limit, independently of the initial condition. This will be confirmed
in the next section by numerical simulations. 

Equations (\ref{prediction1})-(\ref{prediction3}) lead to a certain
number of theoretical predictions. In particular, they imply
that the ratios $\langle\delta N^2(\tau)\rangle/N_H(\tau)$,
$\langle\delta E^2(\tau)\rangle/(N_H(\tau)v_H^4(\tau))$, $\langle\delta
P_i^2(\tau)\rangle/(N_H(\tau)v_H^2(\tau))$ and $\langle\delta N(\tau)\delta
E(\tau)\rangle/(N_H(\tau)v_H^2(\tau))$ reach stationary values at large times.
The approach to these stationary values is exponential in $\tau$, and is
slower for the correlations of the total momentum, since the argument of the
exponential is $p(\zeta_n - \zeta_T)\tau$, while the other quantities evolve
on faster time scales.

\subsection{Numerical simulations}

We now compare our theoretical predictions with the results of Molecular
Dynamics (MD) and Direct Simulation Monte Carlo (DSMC) of a freely evolving
system of $N$ hard disks of diameter $\sigma$ which annihilate with
probability $p$ or collide elastically with probability $1-p$ everytime two
particles meet each other. In the MD case, the particles were localized in a square box
of size $L$ with periodic boundary conditions. The event driven algorithm
\cite{allen} has been used and the initial density has been chosen low enough
to be always in the dilute limit. The parameters for all the MD simulations
were $N(0)=10^5$, $n_H(0)\sigma^2=0.05$, $T_H(0)=1$ and $0<p\le 1$. In
the case of the  
DSMC simulations we have used Bird's algorithm \cite{bird} with the same
values of the parameters, except the density that plays no role. The
  initial velocity distribution is a Maxwellian in both cases. We have
measured the time evolution of the total number of particles and the total
energy, averaging the data over various initial conditions (the total momentum
fluctuates around zero).  We have first checked that the equations
(\ref{evolucion_N})-(\ref{evolucion_E}), with the theoretical predictions
derived in \cite{ptd02} for the cooling rates, correctly describe the decay of
the average global quantities.  In the same way, we have obtained the averaged
values of $N^2(t)$, $E^2(t)$, $P_i^2(t)$ and $N(t)E(t)$ (the correlations
between $P_i$ and $N$ or $E$ are zero).

\begin{figure}
\begin{minipage}[c]{1.0\textwidth}
\begin{center}
\includegraphics[angle=0,width=0.4\textwidth]
{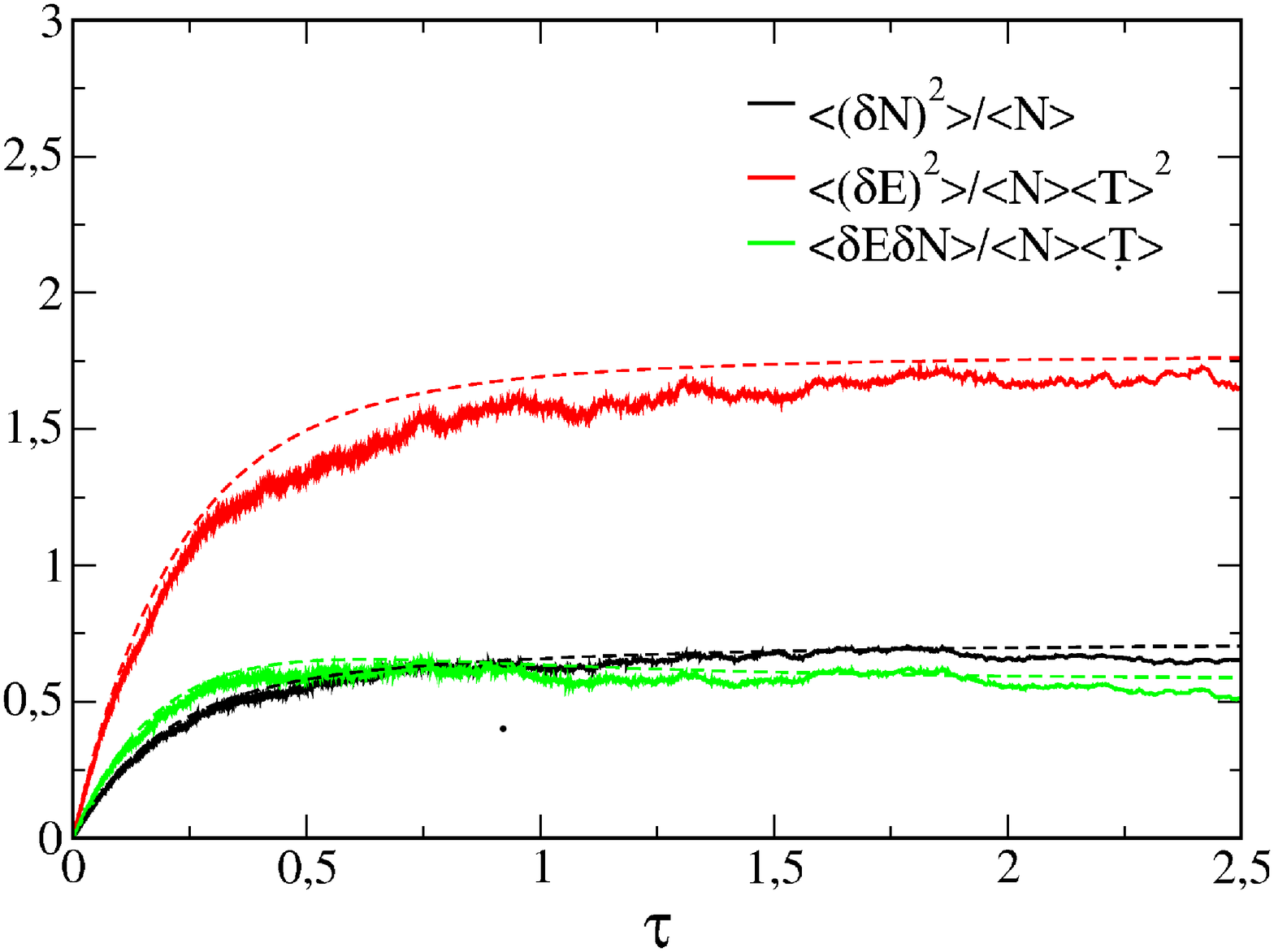}
\includegraphics[angle=0,width=0.4\textwidth]
{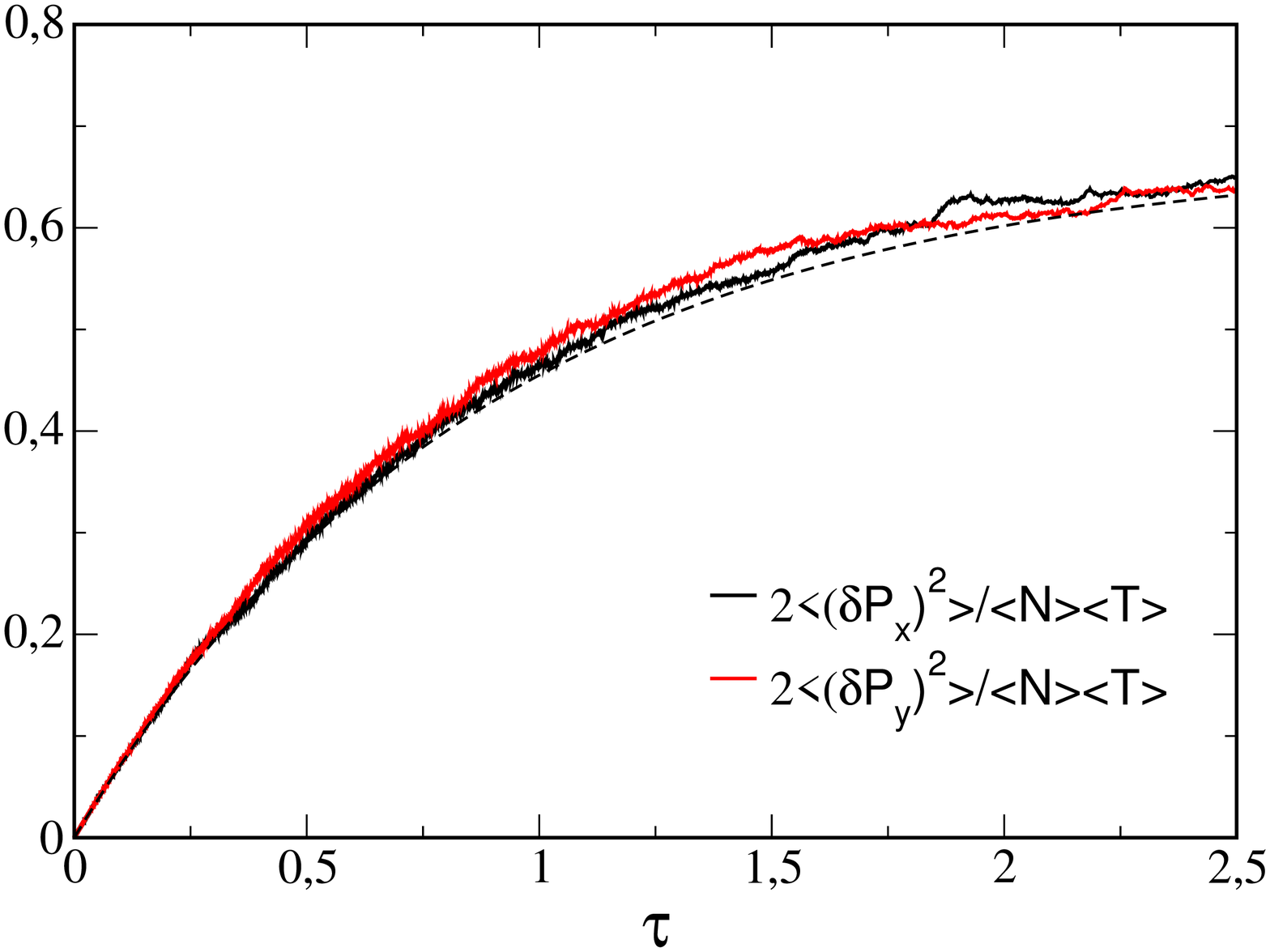}
\end{center}
\end{minipage}
\caption{Scaled second moment of the fluctuations as a function of $\tau$
  for a system with $p=0.8$. These results are from DSMC simulations  and
  have been averaged over 4000 trajectories. }\label{evolucion}
\end{figure}

\begin{figure}
\begin{center}
\includegraphics[angle=0,width=0.4\textwidth]
{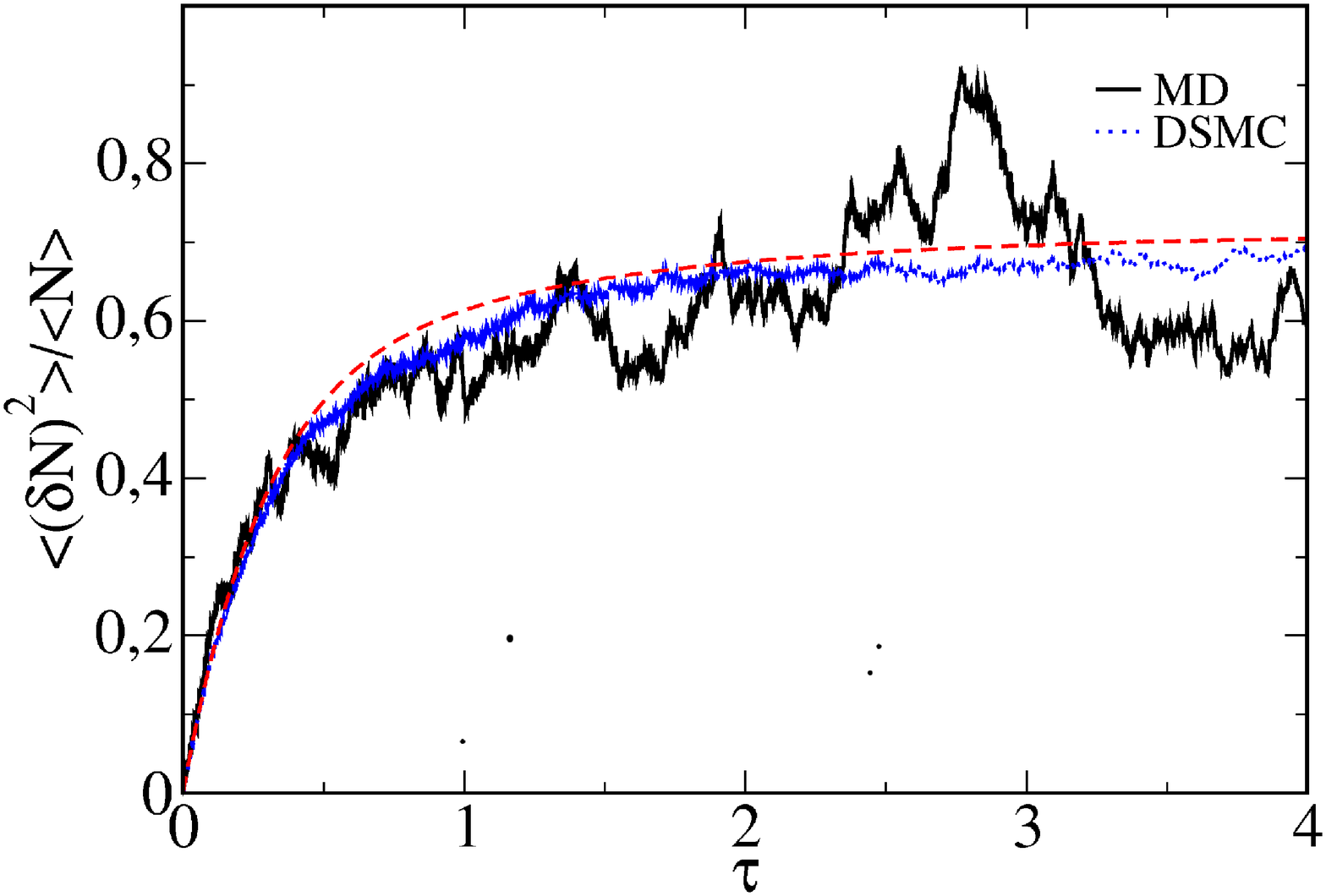}
\includegraphics[angle=0,width=.4\textwidth]
{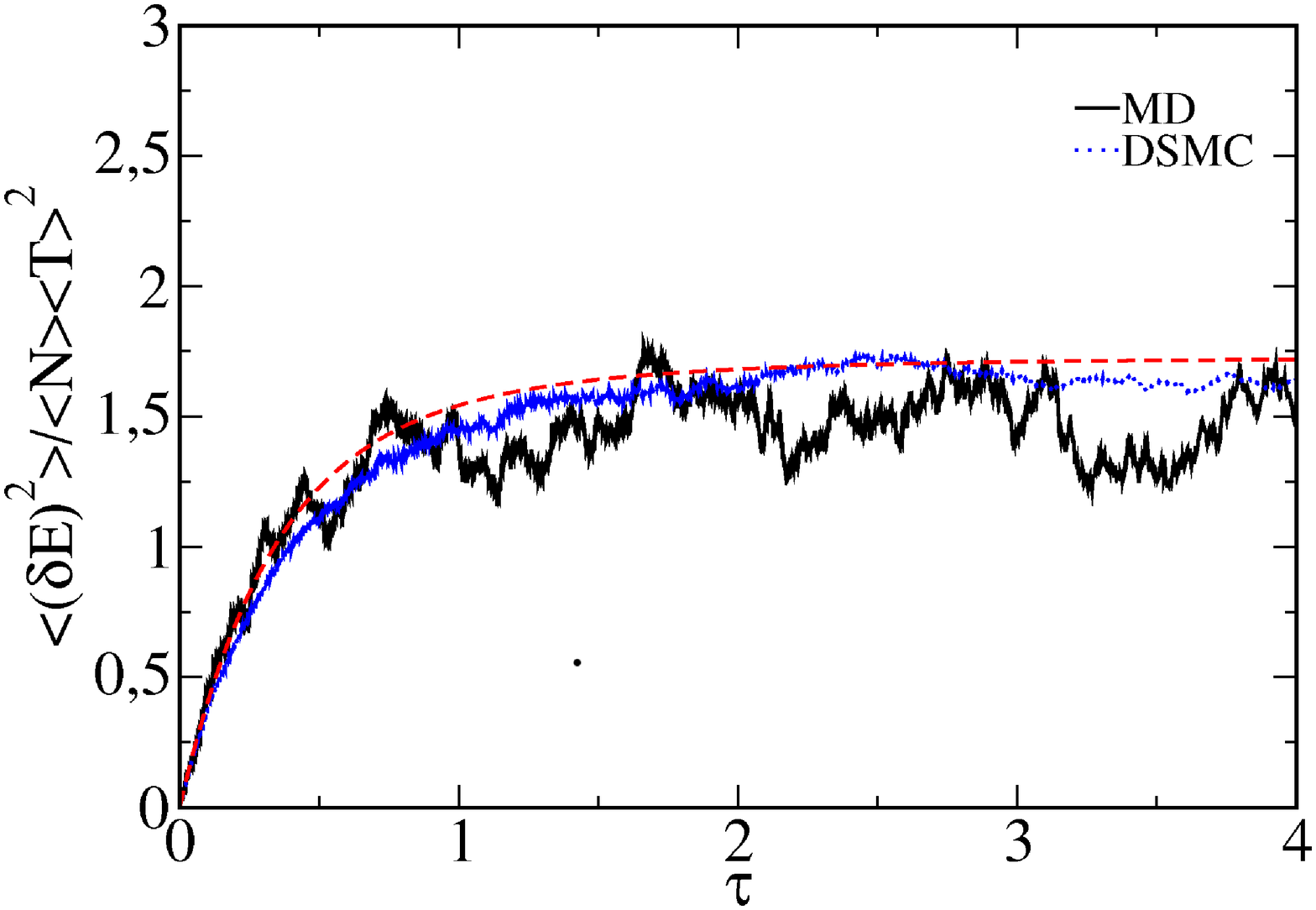}
\end{center}
\caption{Second moment of the fluctuations of the number of particles
  (left panel)
and of the total energy (right panel)
as a 
function of the number of collisions per particle $\tau$, for a system with 
$p=0.5$ and initial number of particles $N=10^5$. The dashed lines are the 
theoretical predictions.}\label{dentau}\label{enertau}
\end{figure}

\begin{figure}
\begin{minipage}[c]{1.0\textwidth}
\begin{center}
\includegraphics[angle=0,width=0.4\textwidth]
{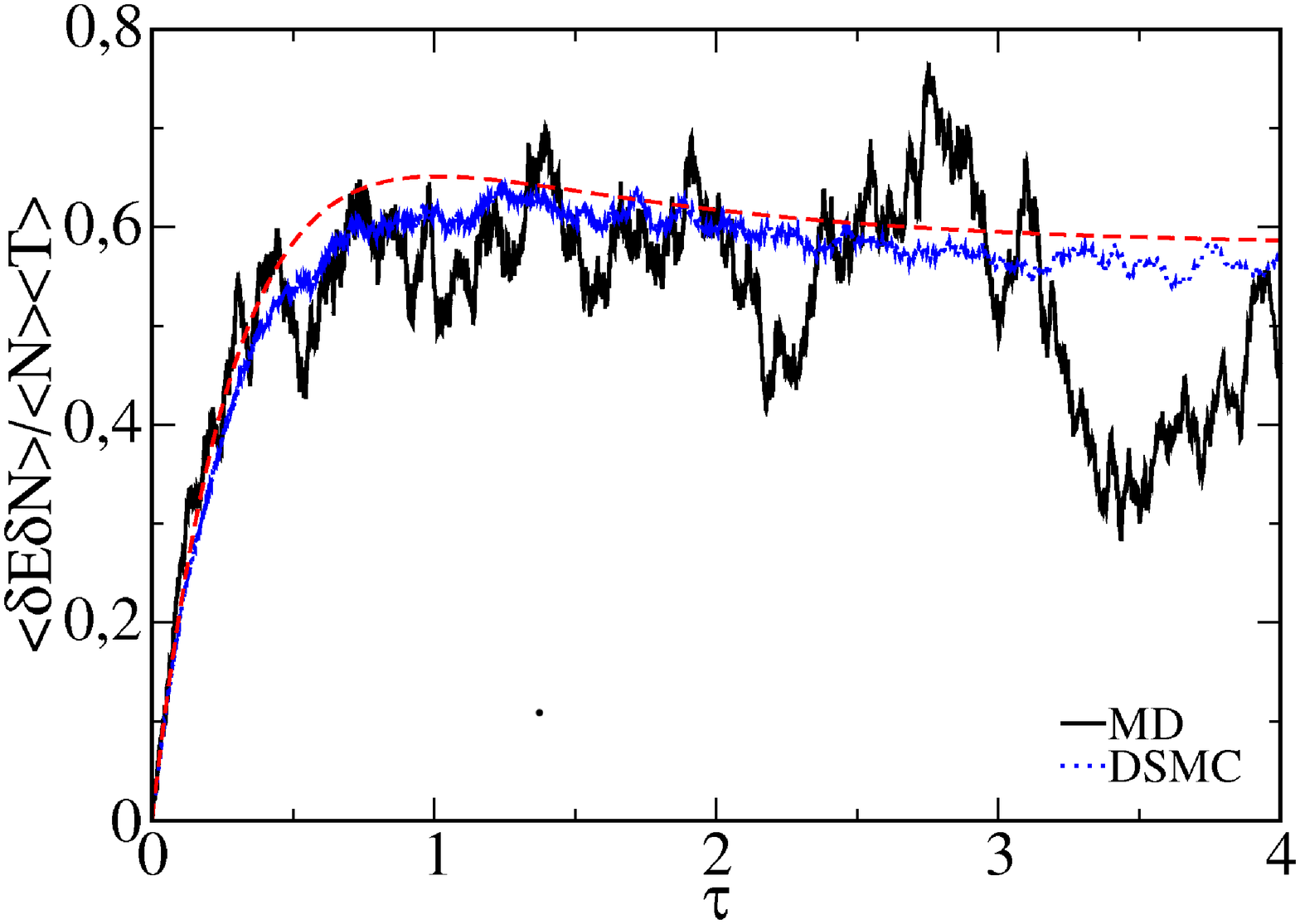}
\includegraphics[angle=0,width=0.4\textwidth]
{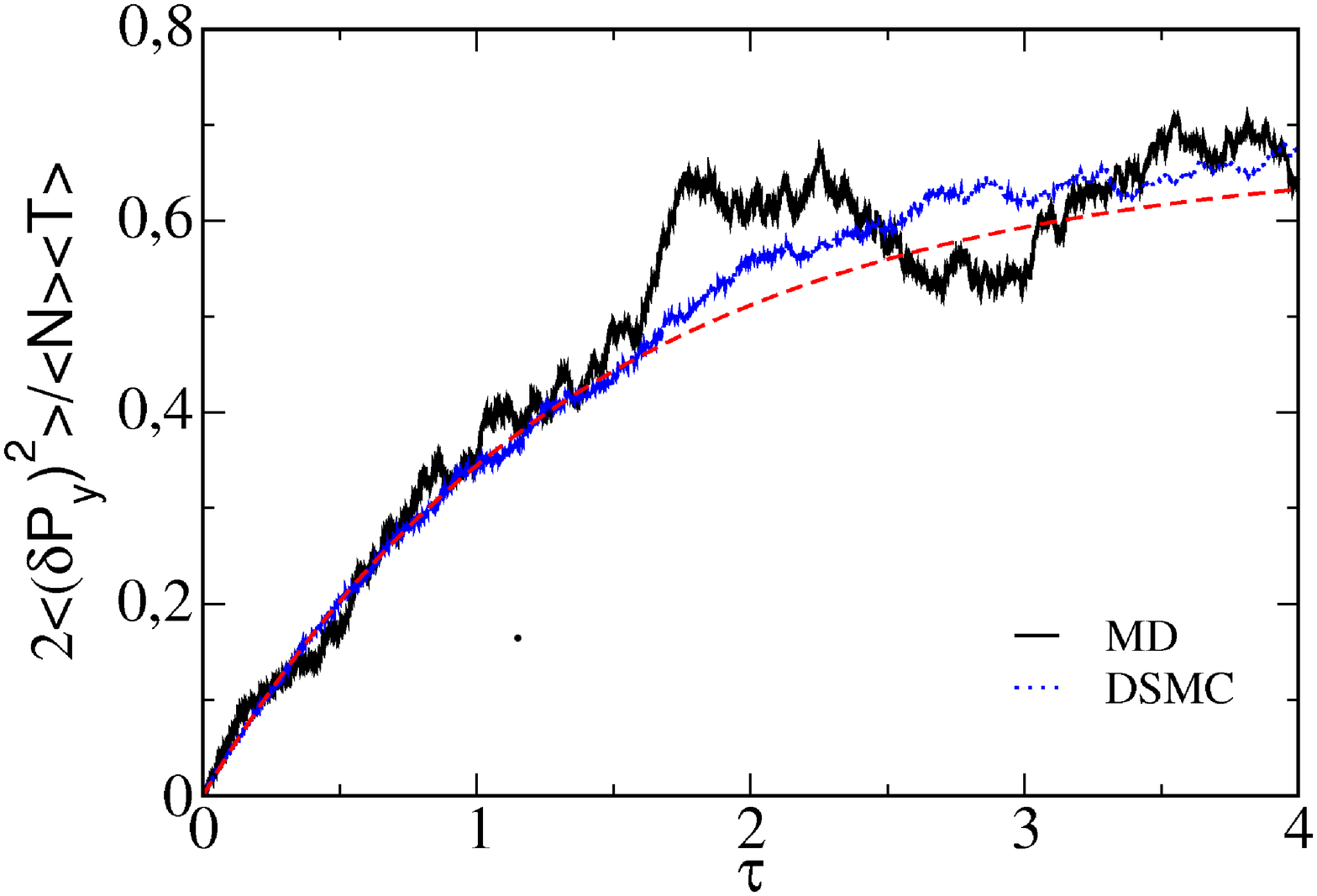}
\end{center}
\end{minipage}
\caption{Correlation between the fluctuations of the total energy and total
  number of particles (left panel), and second moment of the 
fluctuations of the $y$ component of the total
  momentum (right panel),
 as a function of the number of collisions per particle 
$\tau$, for a system with $p=0.5$ and 
initial number of particles $N=10^5$. The dashed lines are the theoretical 
predictions.}\label{ENyPxx}
\end{figure}

Figures \ref{evolucion}, \ref{dentau} and \ref{ENyPxx} show the time evolution
of the various one-time correlation functions considered, for $p=0.5$ and
$p=0.8$. The DSMC results have been averaged over $4000$ trajectories
while the MD 
simulations has been averaged over $150$ trajectories (the DSMC method being
computationally less expensive, it is then possible to average over a larger
number of initial conditions than for the MD simulations).  The dashed lines
are the theoretical predictions, equations 
(\ref{prediction1})-(\ref{prediction3}). Note, however, that the system is not
initially in the HDS : the initial distribution function is a Maxwellian and
not $\chi_H$. Nevertheless, as the difference between these two distributions is
very small (at least for thermal velocities since $a_2\sim 0.1$) and as the
stationary values depend very weakly on $p$, equations 
(\ref{prediction1})-(\ref{prediction3}) predict quite well the time evolution
measured in the simulations.
The ratios $\langle\delta N^2(\tau)\rangle/N_H(\tau)$, $\langle\delta
E^2(\tau)\rangle/(N_H(\tau)v_H^4(\tau))$, and $\langle\delta N(\tau)\delta
E(\tau)\rangle/(N_H(\tau)v_H^2(\tau))$ reach stationary values as predicted.
The fluctuations of the total momentum evolve more slowly, as also predicted,
and the stationary value of $\langle\delta
P_i^2(\tau)\rangle/(N_H(\tau)v_H^2(\tau))$ is barely reached. Note that
$\tau=4$ corresponds for $p=0.5$ to a total number of particles at the end of
the simulation $N\simeq 1700$.

\begin{figure}[h]
\begin{center}
\includegraphics[angle=0,width=0.40\textwidth]
{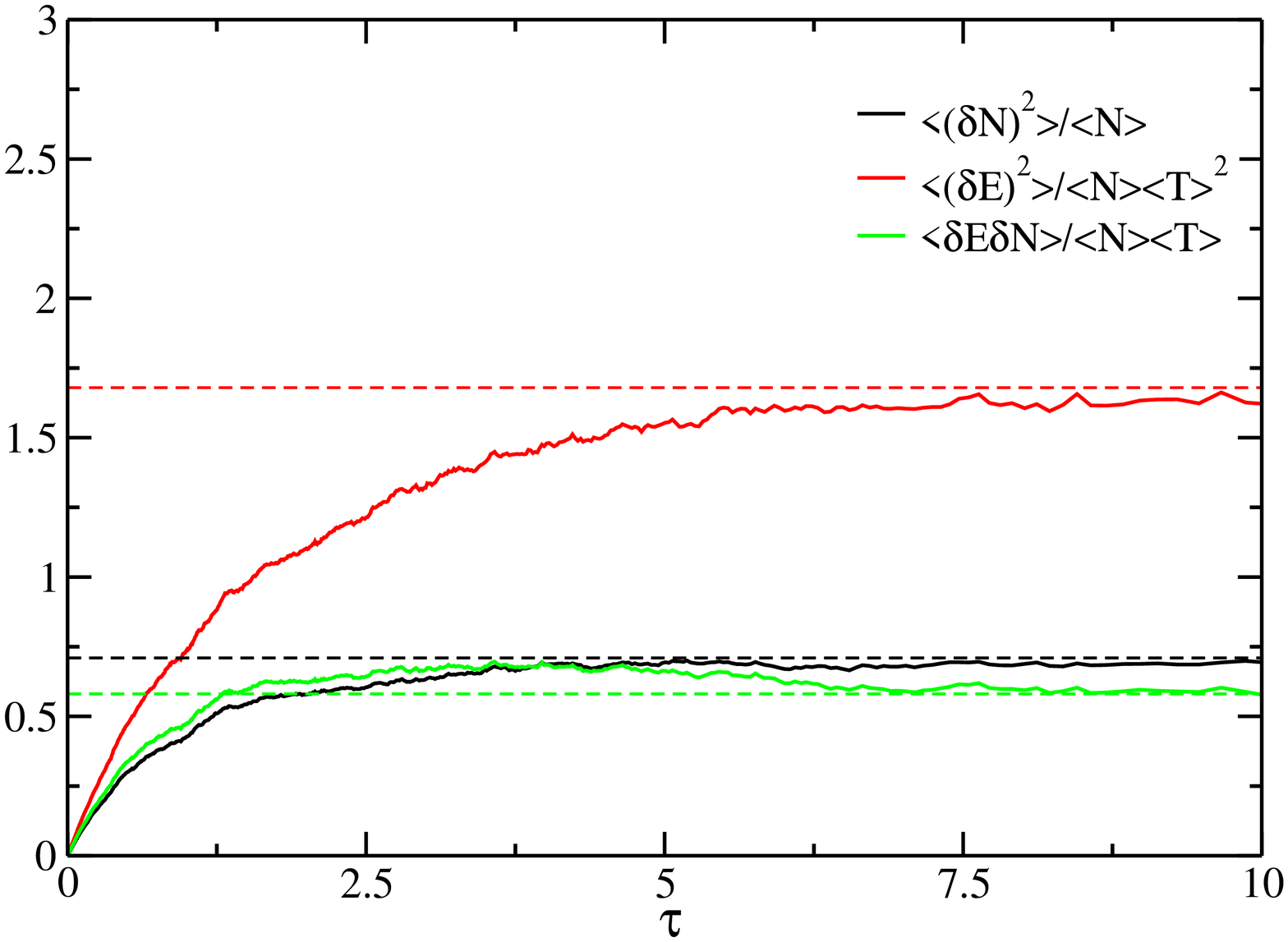}
\includegraphics[angle=0,width=0.40\textwidth]
{fig4B.eps}
\end{center}
\caption{DSMC results for the scaled second moment of the fluctuations
  as a function of $\tau$ for systems with $p=0.2$ (left panel) and
  $p=0.5$ (right panel). The
  dashes lines are the theoretical predictions for the stationary
  values. The initial velocity distribution at $\tau=0$ is uniform
  in a square domain.}\label{fluc_vi_mp_1} 
\end{figure}

\begin{figure}[h]
\begin{minipage}[c]{1.0\textwidth}
\begin{center}
\includegraphics[angle=0,width=0.40\textwidth]
{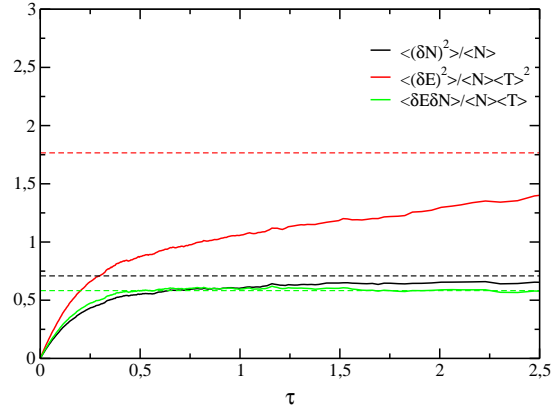}
\end{center}
\end{minipage}
\caption{Same as in Figure \ref{fluc_vi_mp_1} but for a system with $p=0.8$.}\label{fluc_vi_mp_2}
\end{figure}

We have performed simulations starting with other initial conditions 
further from the HDS. The initial velocity distribution function has been 
chosen as a constant function in a square centered in the origin in the
velocity space such that the initial temperature is unity. As seen in 
Figures \ref{fluc_vi_mp_1} and \ref{fluc_vi_mp_2}, we obtain a different
short time evolution but the scaled moments still converge towards the HDS
values, that are represented by the dashed lines. The convergence is 
slower as we increase the value of $p$, and $\langle\delta
E^2(\tau)\rangle/(N_H(\tau)v_H^4(\tau))$, the magnitude which depends
on the higher moments of the velocity distribution, is the most affected.

\begin{figure}[h]
\begin{center}
\includegraphics[angle=0,width=0.4\textwidth]
{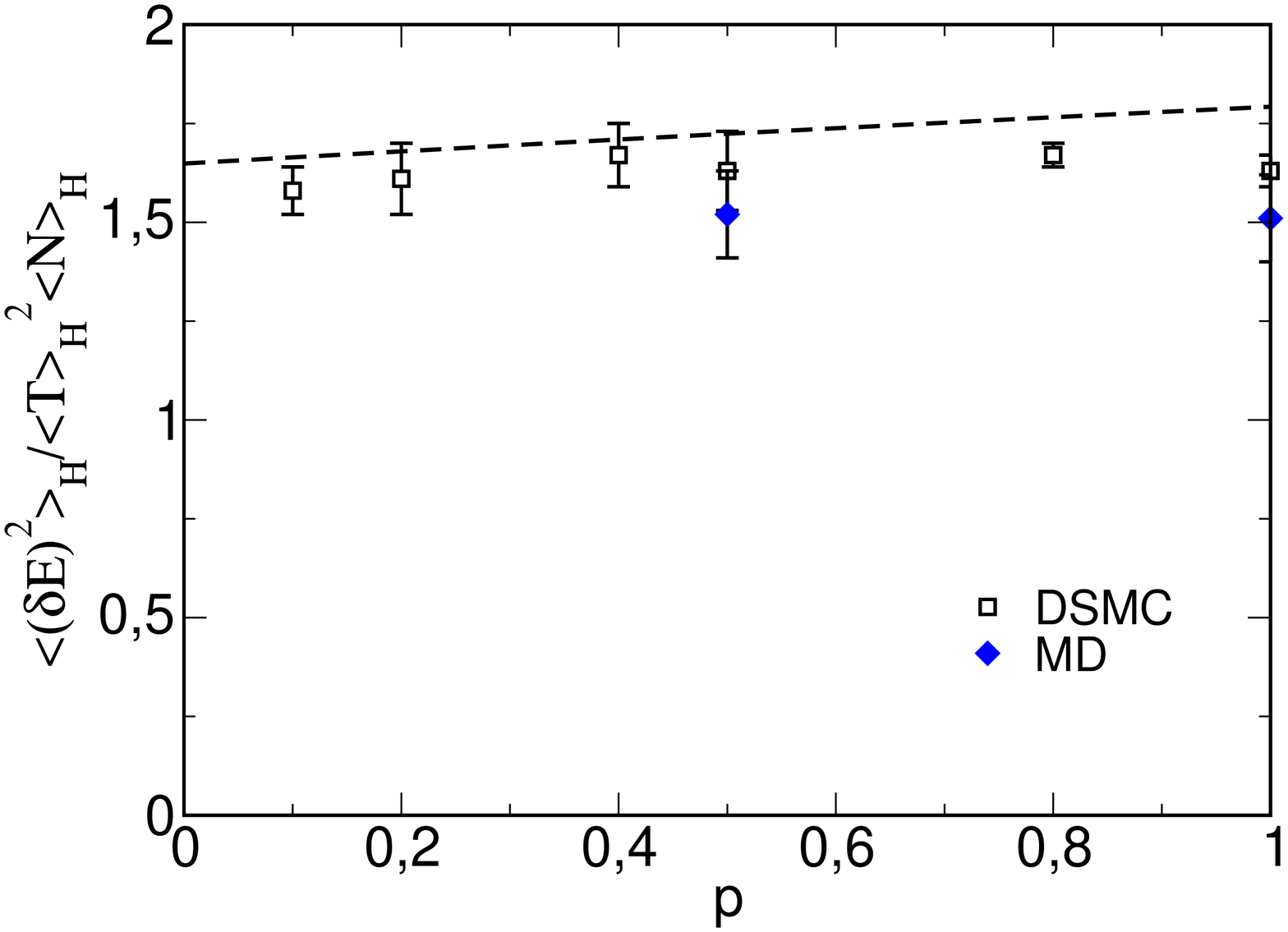}
\includegraphics[angle=0,width=0.4\textwidth]
{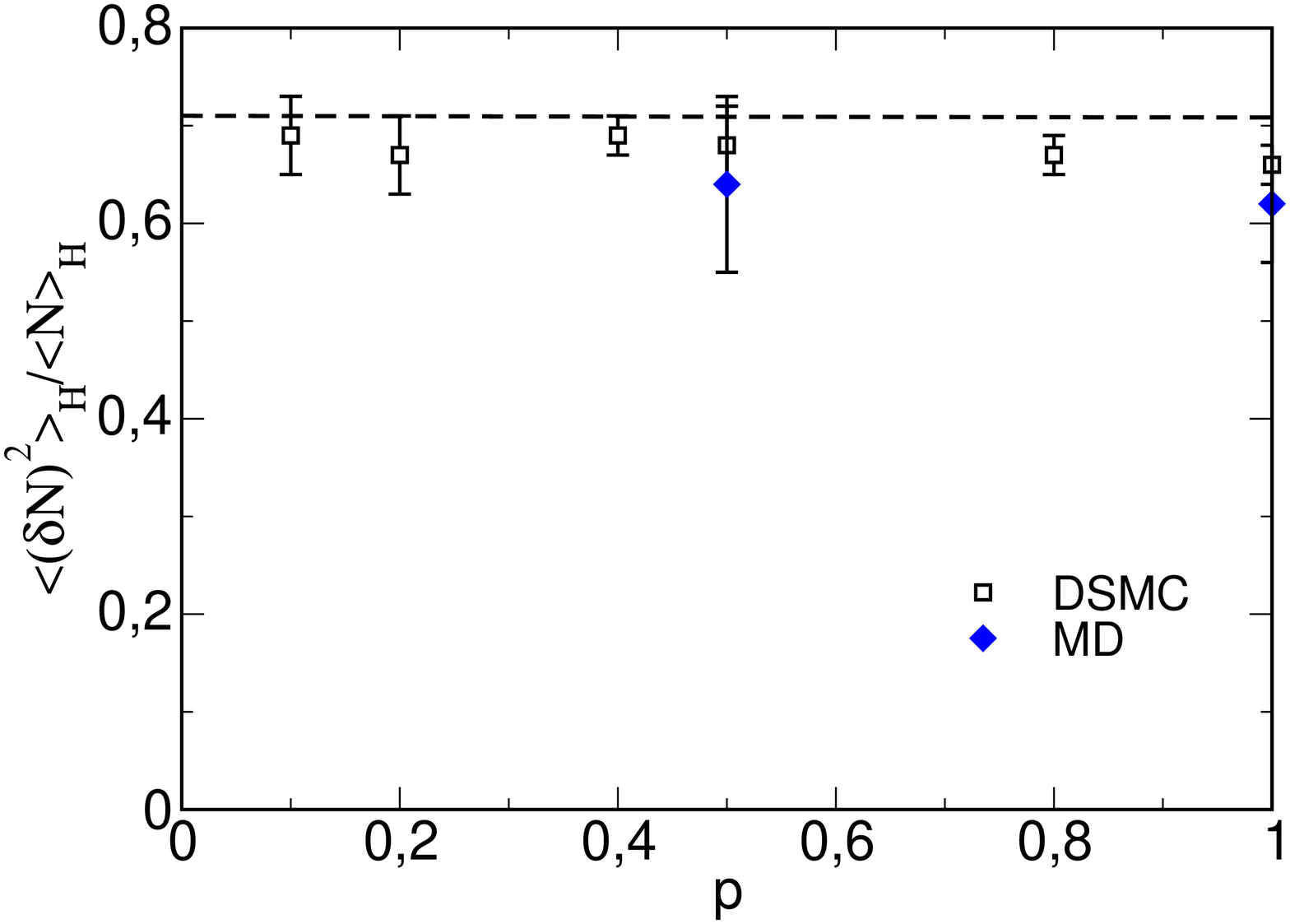}
\end{center}
\caption{Average steady state of the scaled second moment of the
  fluctuations of the number of particles and of the energy fluctuations as
a function of the annihilation probability $p$.
The dashed lines show the large $\tau$ predictions of 
Eqs. (\ref{prediction1}) and (\ref{prediction2}).}\label{stationary1}
\end{figure}

\begin{figure}[h]
\begin{minipage}[c]{1.0\textwidth}
\begin{center}
\includegraphics[angle=0,width=0.4\textwidth]
{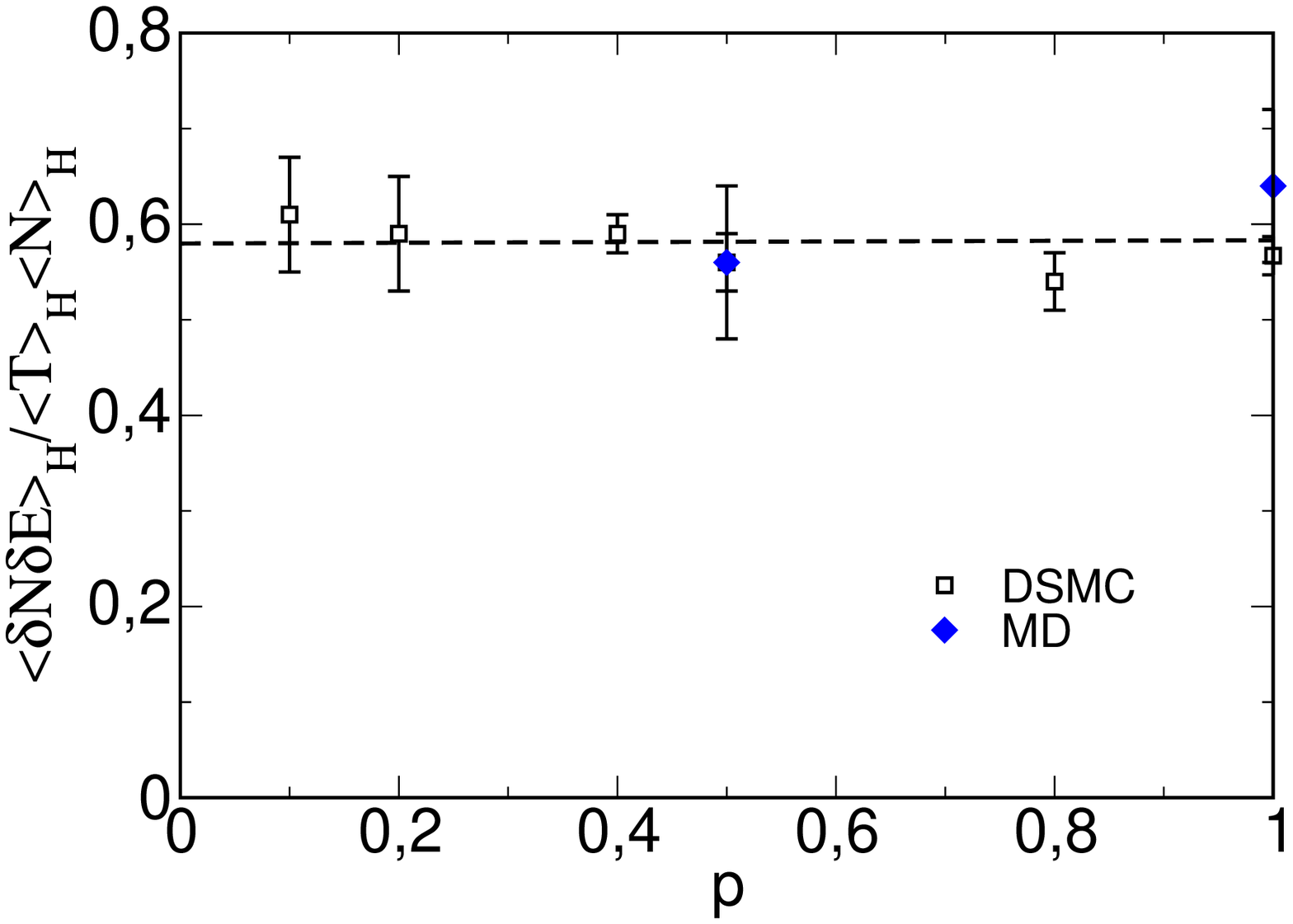}
\end{center}
\end{minipage}
\caption{Average steady values of the one time correlation of the energy and
  number of particles fluctuations as a function of the probability of
  annihilation $p$. The dashed line shows the large $\tau$ prediction of 
Eq. (\ref{prediction3}).}\label{stationary2}
\end{figure}

\begin{figure}[h]
\begin{minipage}[c]{1.0\textwidth}
\begin{center}
\includegraphics[angle=0,width=.4\textwidth]{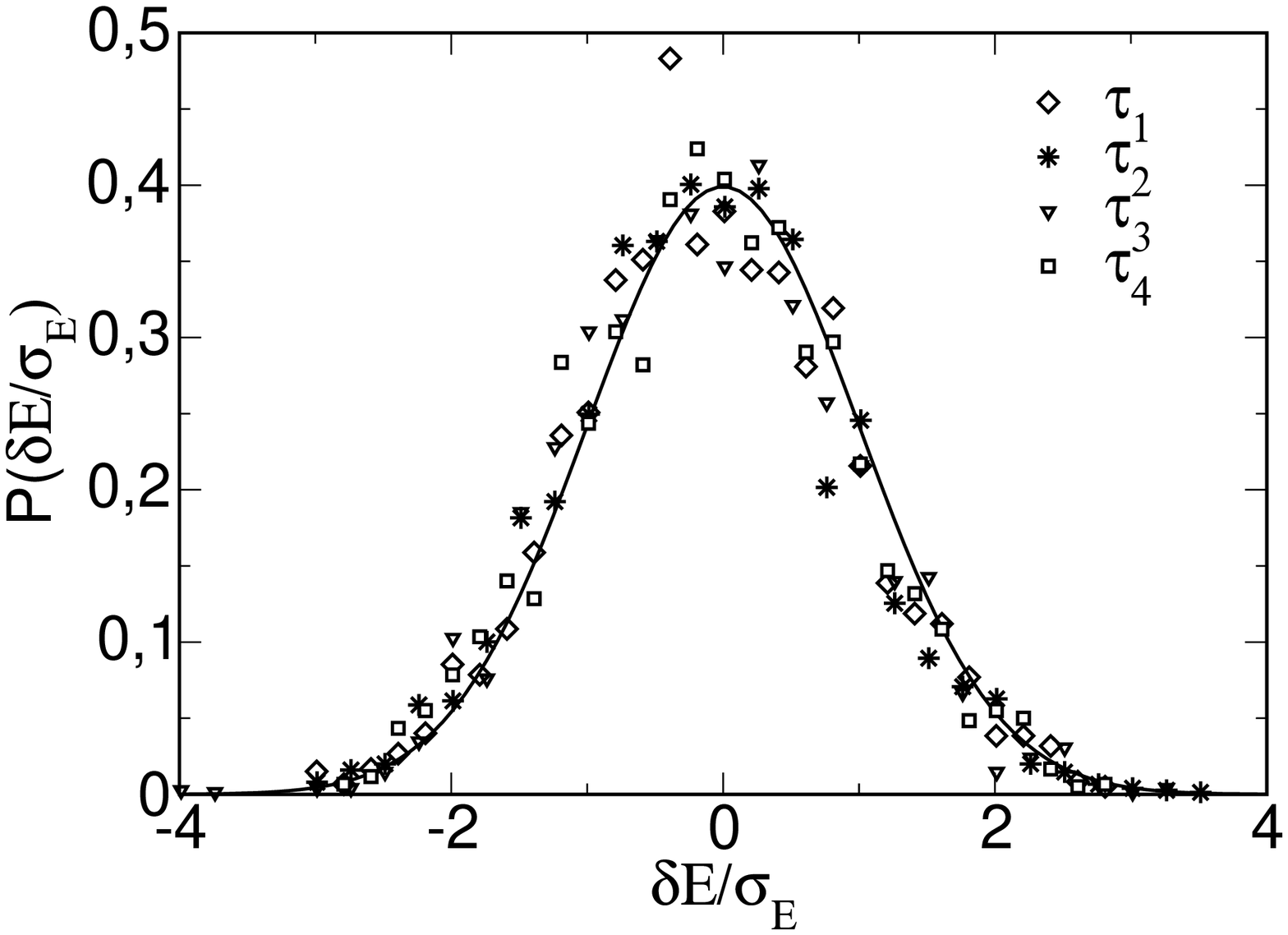}
\includegraphics[angle=0,width=.4\textwidth]{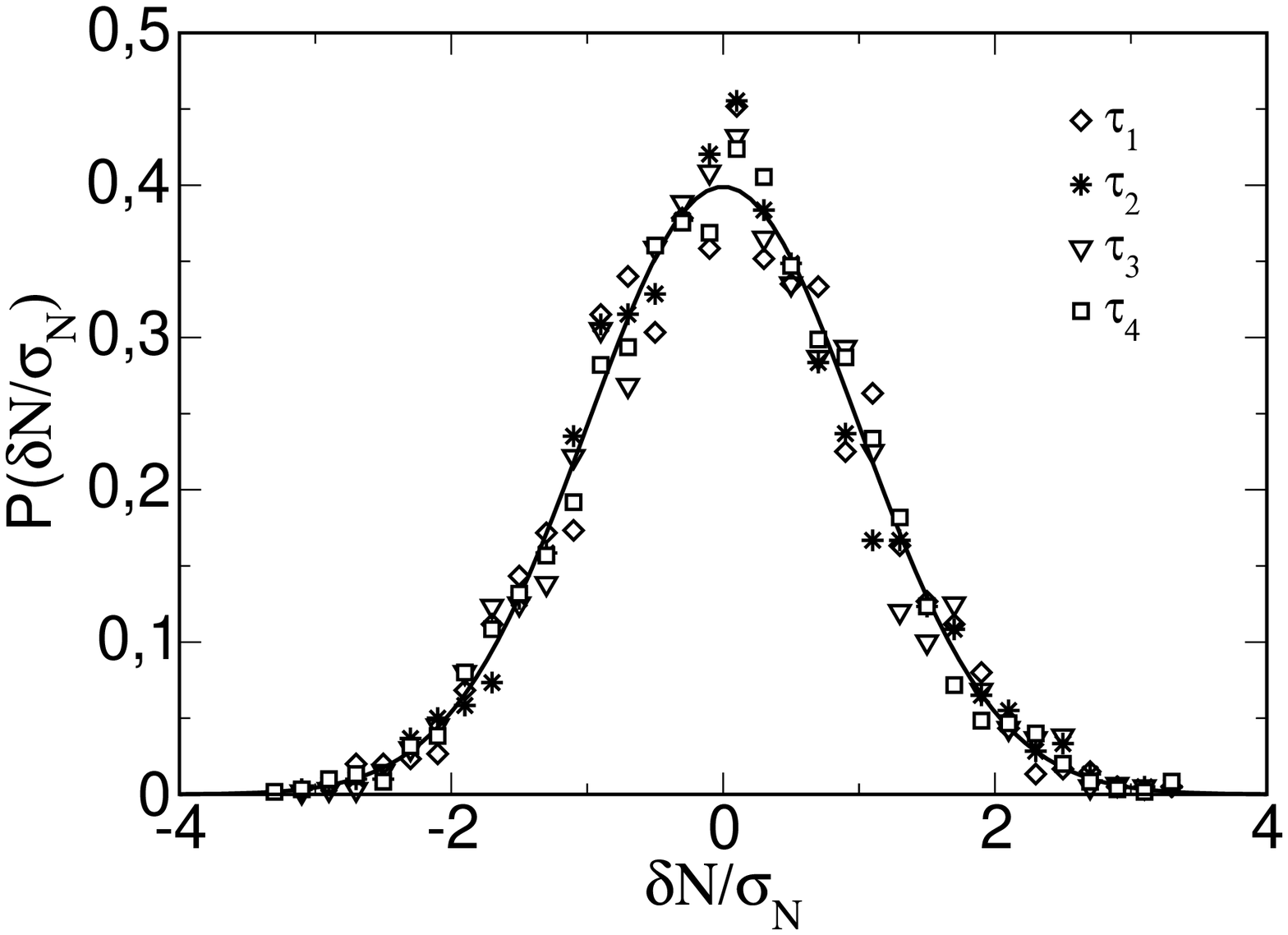}
\end{center}
\end{minipage}
\caption{Normalized distribution of the relative energy (left panel)
  and number of 
  particles fluctuations (right panel) for a system with $p=0.5$. The
  symbols are from 
  DSMC simulations and for four different values of $\tau$, $\tau_1=0.69$, $\tau_2=0.98$, $\tau_3=1.37$ and $\tau_4=2.05$. The solid line
  is a Gaussian with unit variance.}\label{histogramas}
\end{figure}

Figures  \ref{stationary1} and \ref{stationary2} show the comparison
between the stationary values of the various ratios measured
in the simulations and the theoretical predictions in Eq. 
(\ref{prediction1}-\ref{prediction3}) at large $\tau$. The agreement
is very good for all values investigated. 
We have also computed the probability distribution for the
number of particles, energy and momentum. 
As we can see in Fig. \ref{histogramas}, where we have considered
a system with $p=0.5$, they are correctly described 
by a Gaussian distribution. The figure displays the distribution
at four different times, showing that the
shape of the probability distributions does not vary during the dynamical
evolution. Similar results have been obtained for the
probability distribution of the total momentum.

\section{Two-time correlation function in the HDS}
\label{Sec. 6}

In this section, we study the two-time correlation function of the global 
quantities in the HDS. To this aim, we consider two dynamical 
variables $A(t)$ and $B(t)$ as in (\ref{iv.4}), and compute
the correlations 
$\langle \delta A(t)\delta B(t^\prime)\rangle_H$ for $t>t'$, which
are obtained from $h_{1,1,H}$ through equation (\ref{eq2times}).

As in the previous section, we start by integrating out the 
spatial degrees of freedom and consider
\begin{equation}
\psi_H(\tau,\tau^\prime,\mathbf{c}_1,\mathbf{c}_2)\equiv\int d\mathbf{r}_{12}
\tilde{h}_H(\mathbf{r}_{12};\mathbf{c}_1,\tau;\mathbf{c}_2,\tau^\prime),
\end{equation}
whose evolution equation is obtained by integrating (\ref{ec.ev.h11}) over space variables
\begin{equation}
\frac{\partial}{\partial\tau}\psi_H(\tau,\tau^\prime,\mathbf{c}_1,\mathbf{c}_2)
=\Lambda(\mathbf{c}_1)\psi_H(\tau,\tau^\prime,\mathbf{c}_1,\mathbf{c}_2).
\end{equation}
This equation has to be solved with the initial condition
\begin{equation}
\psi_H(\tau',\tau',\mathbf{c}_1,\mathbf{c}_2)
=\chi_H(\mathbf{c}_1)\delta(\mathbf{c}_1-\mathbf{c}_2)
+\phi_{H}(\tau',\mathbf{c}_1,\mathbf{c}_2),
\end{equation}
where we have taken into account the scaling of $f_H$ (\ref{iii.1})
and $g_{2,H}$ (\ref{iii.11}). Then, using the approximation 
(\ref{aproximacionPL}), we obtain
\begin{eqnarray}
P_{12}\psi_H(\tau,\tau^\prime,\mathbf{c}_1,\mathbf{c}_2)&=&
e^{\Lambda(\mathbf{c}_1)(\tau-\tau^\prime)}P_{12}
\psi_H(\tau',\tau',\mathbf{c}_1,\mathbf{c}_2)\nonumber\\
&=&\langle\bar{\xi}_1(\mathbf{c}_1)\vert
\psi_H(\tau',\tau',\mathbf{c}_1,\mathbf{c}_2)
\rangle\xi_1(\mathbf{c}_1)\nonumber\\
&+&\langle\bar{\xi}_2(\mathbf{c}_1)\vert
\psi_H(\tau',\tau',\mathbf{c}_1,\mathbf{c}_2)
\rangle\xi_2(\mathbf{c}_1)e^{-p(\zeta_T+2\zeta_n)(\tau-\tau^\prime)}\nonumber\\
&+&\sum_i\langle\bar{\xi}_{3i}(\mathbf{c}_1)\vert
\psi_H(\tau',\tau',\mathbf{c}_1,\mathbf{c}_2)\rangle
\xi_{3i}(\mathbf{c}_1)e^{p\zeta_T(\tau-\tau^\prime)}.
\end{eqnarray}
In the large time limit $\tau, \tau' \to \infty$, $\tau-\tau'$
finite (and positive) we can replace
$\psi_H(\tau',\tau',\mathbf{c}_1,\mathbf{c}_2)$ by 
$\chi_H(\mathbf{c}_1)\delta(\mathbf{c}_1-\mathbf{c}_2)
+\phi^s_{H}(\mathbf{c}_1,\mathbf{c}_2)$, so that
\begin{eqnarray}
\langle\delta N(\tau)\delta N(\tau^\prime)\rangle_H&=&N_H(\tau)
\left\{-A_{1,1}-zA_{1,2}-zA_{1,2}e^{-p(\zeta_T+2\zeta_n)(\tau-\tau^\prime)}\right.
\nonumber\\
&-&\left.z^2A_{2,2}e^{-p(\zeta_T+2\zeta_n)(\tau-\tau^\prime)}\right\},\\
\langle\delta P_i(\tau)\delta P_j(\tau^\prime)\rangle_H&=&
\delta_{ij}N_H(\tau)v_H(\tau)v_H(\tau^\prime)\left[a^s_{3i,3i}+\frac{1}{2}\right]
e^{p\zeta_T(\tau-\tau^\prime)},\label{vi.18}\\
\langle\delta E(\tau)\delta E(\tau^\prime)\rangle_H&=&
\left(\frac{dm}{4}\right)^2N_H(\tau)v_H^2(\tau)v_H^2(\tau^\prime)
\left\{-A_{1,1}+(z+2)A_{1,2}\right.\nonumber\\
&+&\left.\left[-(z+2)^2A_{2,2}+(z+2)A_{1,2}\right]
e^{-p(\zeta_T+2\zeta_n)(\tau-\tau^\prime)}\right\},\\
\langle\delta N(\tau)\delta E(\tau^\prime)\rangle_H&=&m\frac{d}{4}N_H(\tau)
v_H^2(\tau^\prime)\left\{A_{1,1}-(z+2)A_{1,2}\right.\nonumber\\
&+&\left.\left[-z(z+2)A_{2,2}+zA_{1,2}\right]
e^{-p(\zeta_T+2\zeta_n)(\tau-\tau^\prime)}\right\} .
\end{eqnarray}

\begin{figure}
\begin{minipage}[c]{1.0\textwidth}
\begin{center}
\includegraphics[angle=0,width=0.4\textwidth]
{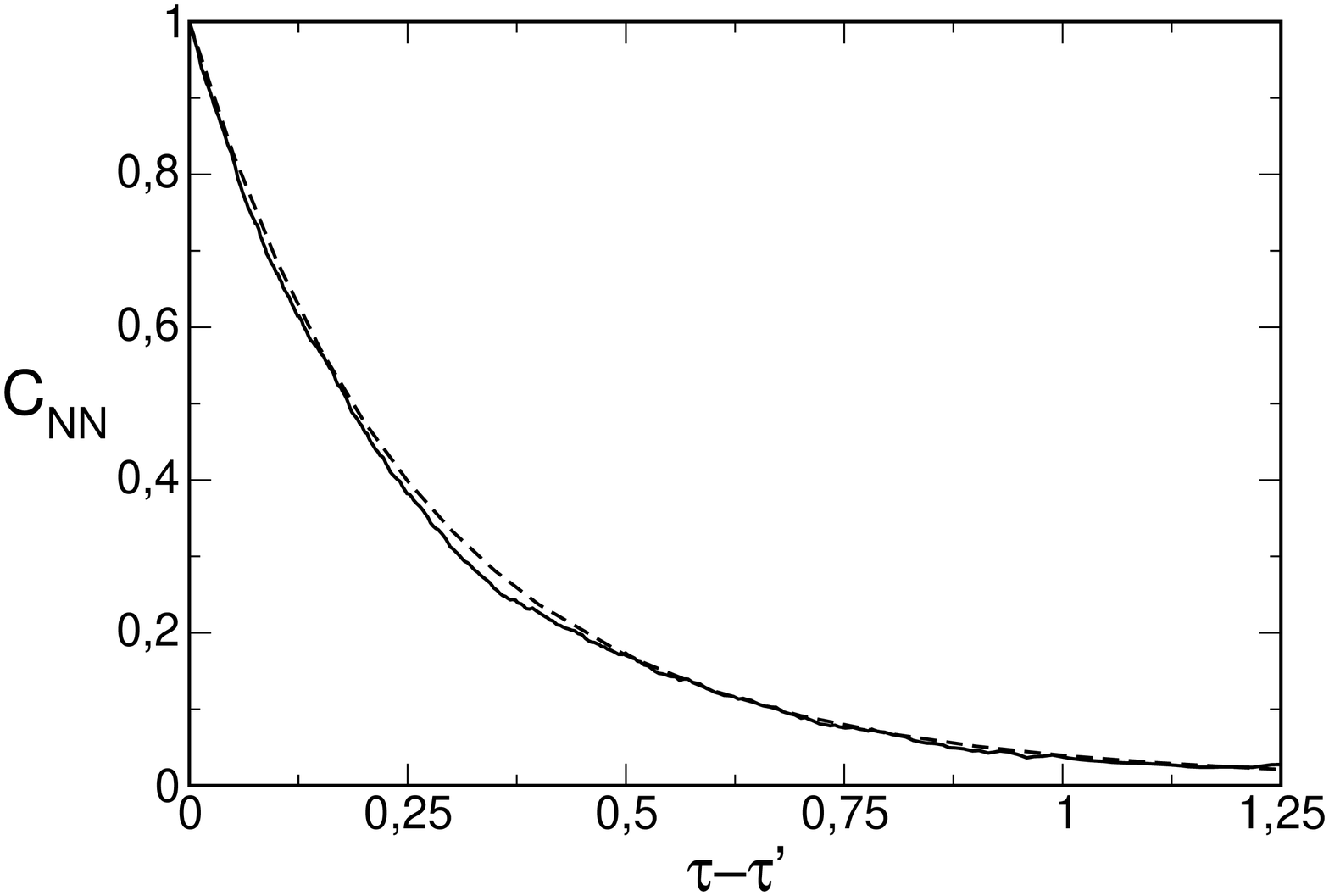}
\includegraphics[angle=0,width=0.4\textwidth]
{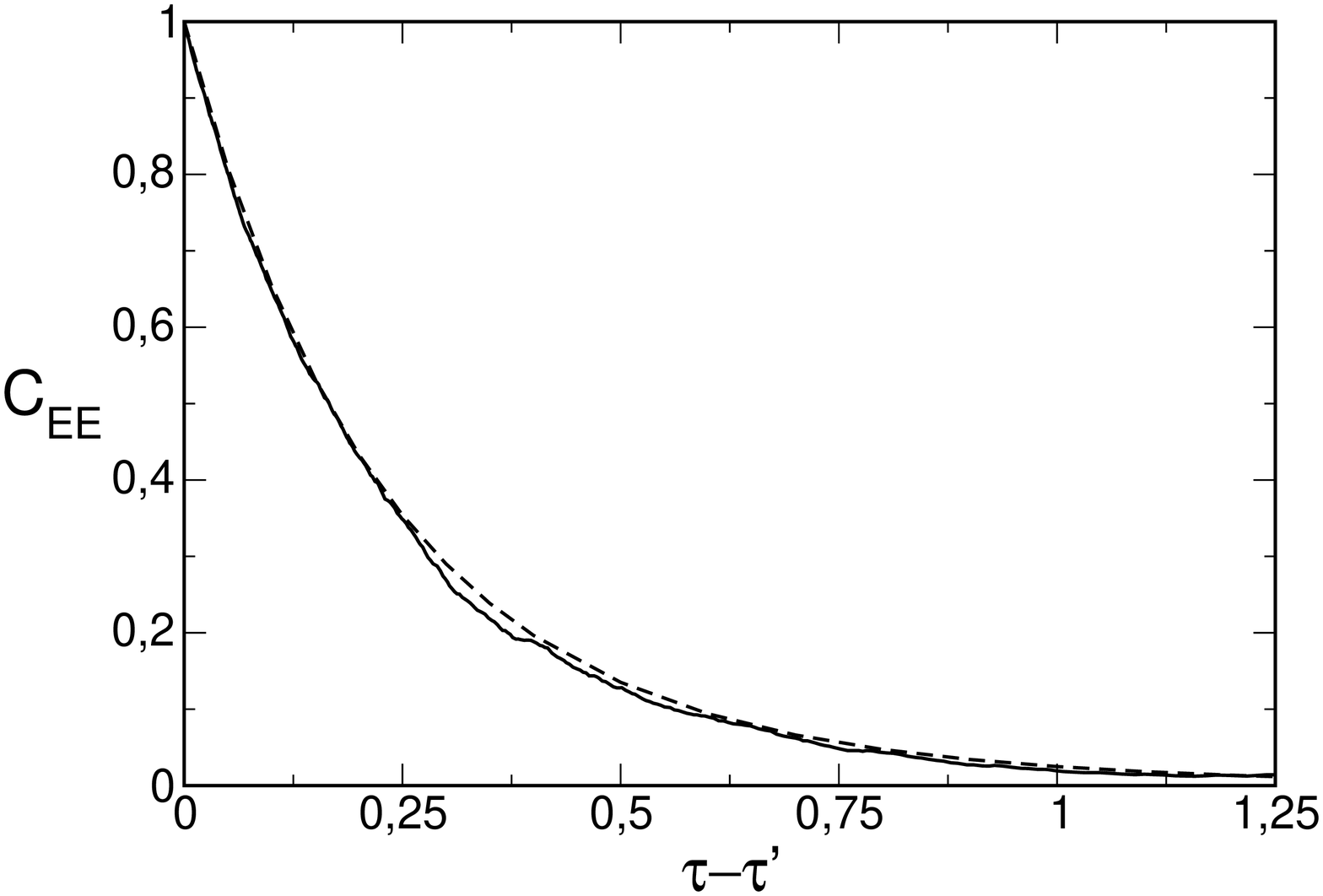}
\end{center}
\end{minipage}
\caption{Decay of the two-time correlation of the number of particles, 
$C_{NN}$, and the energy, $C_{EE}$, for a system with $p=1$, measured with
DSMC simulations. The dashed line
is the theoretical prediction.}\label{decaimientoNyE}
\end{figure}

\begin{figure}
\begin{minipage}[c]{1.\textwidth}
\begin{center}
\includegraphics[angle=0,width=.4\textwidth]
{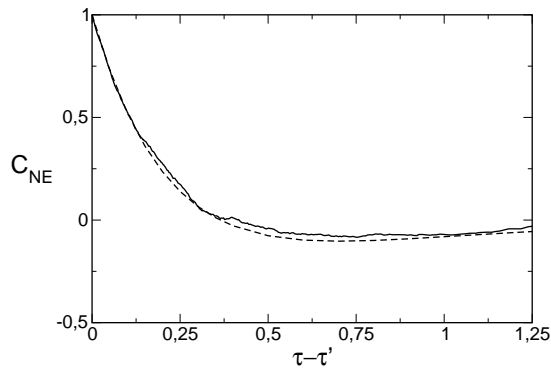}
\end{center}
\end{minipage}
\caption{Decay of the two-time correlation of the number of particles and the 
energy, $C_{NE}$ for a system with $p=1$. The dashed line is the theoretical 
prediction.}\label{decaimientoNE}
\end{figure}
\begin{figure}
\begin{minipage}[c]{1.\textwidth}
\begin{center}
\includegraphics[angle=0,width=.4\textwidth]
{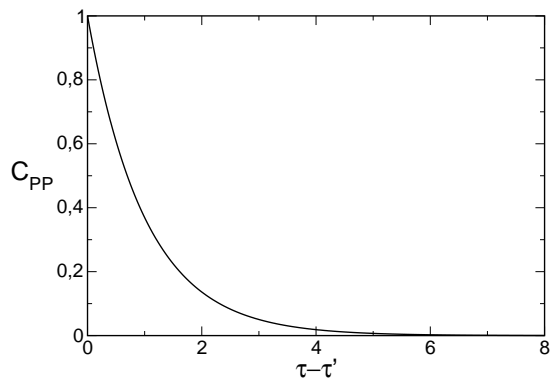}
\end{center}
\end{minipage}
\caption{Theoretical prediction for the decay of the
momentum correlation function  $C_{PP}(\tau,\tau^\prime)$, defined in the
main text,  as a function of
$\tau-\tau^\prime$ for a system with 
$p=0.5$.}\label{decaimientoP}
\end{figure}

In the $\tau$ scale, it can be seen from 
equations (\ref{evolucion_N}) and (\ref{evolucion_E}) that $N_H$ and
$v_H$ decay exponentially. For $A,B=N,E,P$, the 
normalized correlation functions
\begin{equation}
C_{AB}(\tau,\tau^\prime)
=\langle\delta A(\tau)\delta B(\tau^\prime)\rangle/\langle\delta
A(\tau^\prime)\delta B(\tau^\prime)\rangle ,
\end{equation}
become therefore time-translation invariant, {\it i.e.} functions of
$\tau-\tau'$ once the stationary regime for the ratios such as
$\langle\delta N^2(\tau)\rangle/N_H(\tau)$ has been reached (see previous
section). We have checked numerically that this is indeed the case, and
we compare in Fig. \ref{decaimientoNyE} and \ref{decaimientoNE},
the evolution of $C_{NN}(\tau-\tau')$, $C_{EE}(\tau-\tau')$
and $C_{NE}(\tau-\tau')$ measured in DSMC simulations
(for $p=1$, averaged over $4000$ trajectories) 
with the theoretical predictions. The agreement is 
very good. Figure
\ref{decaimientoP} also shows the theoretical prediction for the decay of the
momentum correlation function $C_{PP}(\tau,\tau^\prime)= \langle\delta
P_i(\tau)\delta P_j(\tau^\prime)\rangle/ \langle\delta P_i(\tau^\prime)\delta
P_j(\tau^\prime)\rangle$ for $p=0.5$.  The characteristic decay time of
$C_{PP}$ is of the order of $\tau\sim 4$.  Because the time to reach the
stationary regime for $\langle\delta
P_i^2(\tau)\rangle/(N_H(\tau)v_H^2(\tau))$ is $\tau\sim 4$, as shown in Fig.
\ref{ENyPxx}, we would need to reach $\tau\simeq 8$ in the numerical
simulations in order to display numerical results for $C_{PP}$, which means
that we would need to consider simulations with an initial
number of particles of the order of $N(0)\sim 10^7$.

\section{Conclusions}\label{Sec. 8}

In this paper, we have formulated a general theory for fluctuations and
correlations in a dilute probabilistic ballistic annihilation system.
The theory
has been particularized to the homogeneous decay state, taking advantage
of its scaling properties. For this state we have focused on the study of the
fluctuations of the total number of particles, total momentum and total
energy, evaluating the two-time correlation functions between these quantities
in the hydrodynamic approximation. The fluctuations of the total number of
particles, total momentum and total energy, once conveniently rescaled,
converge to stationary values. The convergence is exponential in the natural
time-scale $\tau$ given by the number of collisions, and the corresponding
rates are simple combinations of the cooling rates. The stationary values are
obtained as functionals of the distribution function and can be computed in
the first Sonine approximation. We have also obtained theoretical expressions 
for the two-time correlations of global observables.  All our theoretical
predictions have been successfully compared with the results of Molecular
Dynamics and DSMC numerical simulations, providing strong
support for the theory developed here, including the hydrodynamic description 
in terms of the lowest order eigenfunctions and eigenvalues of the linearized
Boltzmann collision operator.

As a side-remark, we note that
the correlation functions contain two parts: one coming from the one-particle
distribution function, and another one that comes from velocity
correlations. Nevertheless, it must be stressed that the existence of these
velocity correlations does not imply a violation of the ``molecular chaos''
assumption that underlies the Boltzmann equation. This is because the latter
only refers to the precollisional part of the two-body distribution function 
(at contact).
 
The fact that the two-time correlation functions decay on a time-scale
determined by the cooling rates reflects the intuitive notion that their
dynamic is essentially of macroscopic character, compatible with Onsager's
regression hypothesis (see e.g. \cite{chandler}). 
To analyze this point in a deeper way, we show in
Appendix \ref{Sec. 7} that a description of the fluctuations 
$\delta N$, $\delta E$ and $\delta P$ in terms of
linear Langevin equations can be obtained, using for the deterministic part
the evolution equations for a linear perturbation around the HDS. The
(Gaussian and delta correlated in time) noise terms in the Langevin 
equations can then be adjusted in order to obtain the same amplitudes for 
the one-time correlation
functions as with our theory. In this respect, the results
derived in Appendix \ref{Sec. 7} may be envisioned
as formulating a fluctuation-dissipation theorem for the
homogeneous decay state under scrutiny in this paper.
The amplitudes of the noise terms are however complicated functions
of moments of the one-particle distribution functions, and are
not clearly related to macroscopic quantities such as the cooling rates. 

\begin{acknowledgments}
We would like to thank the Agence Nationale de la Recherche
for financial support. M.~I.~G.~S.~acknowledges financial support from Becas de
la 
Fundaci\'on La Caixa y el Gobierno Franc\'es and from the HPC-EUROPA project (RII3-CT-2003-506079),
with the support of the European Community Research Infrastructure Action.

\end{acknowledgments}

\appendix
\section{Expressions for $a^s_{\beta_1,\beta_2}$}\label{appendixA}
In this Appendix we compute the expressions for the coefficients
$a^s_{\beta_1,\beta_2}$. Applying the projector $P_{12}$ to (\ref{iv.2}) yields, under the
assumption $P_{12}\Lambda(\mathbf{c}_i)=P_{12}\Lambda(\mathbf{c}_i)P_{12}$,
\begin{equation}\label{eq.phiHh}
\left[\Lambda(\mathbf{c}_1)+\Lambda(\mathbf{c}_2)-2p\zeta_n \right]
\phi_H^{s\,(h)}(\mathbf{c}_1,\mathbf{c}_2)
=-\gamma P_{12} T(\mathbf{c}_1,\mathbf{c}_2)
\chi_H(\mathbf{c}_1)\chi_H(\mathbf{c}_2).
\end{equation}
Using the definition (\ref{b.15}) of the coefficients $a^s_{\beta_1,\beta_2}$, we
then obtain the set of equations
\begin{equation}\label{b.16}
\sum_{\beta_1}^3\sum_{\beta_2}^3a^s_{\beta_1,\beta_2}
(\lambda_{\beta_1}+\lambda_{\beta_2}-2 p \zeta_n )
\xi_{\beta_1}(\mathbf{c}_1)\xi_{\beta_2}(\mathbf{c}_2)
=-P_{12}\gamma T(\mathbf{c}_1,\mathbf{c}_2)
\chi_H(\mathbf{c}_1)\chi_H(\mathbf{c}_2),
\end{equation}
and it is straightforward to write
\begin{equation}\label{a_beta1beta2}
a^s_{\beta_1,\beta_2}=
-\frac{\langle\bar{\xi}_{\beta_1}(\mathbf{c}_1)\bar{\xi}_{\beta_2}
(\mathbf{c}_2)\vert\gamma T(\mathbf{c}_1,\mathbf{c}_2)
\chi_H(\mathbf{c}_1)\chi_H(\mathbf{c}_2)\rangle}
{\lambda_{\beta_1}+\lambda_{\beta_2}-2p\zeta_n }.
\end{equation}

Given the expression of the functions 
$\{\bar{\xi}_i(\mathbf{c})\}_{i=1}^3$, 
and taking into account the relations
\begin{eqnarray}
p\zeta_n&=&-\frac{\gamma}{2}\int\!\!d\mathbf{c}_1\!\!\int\!\!d\mathbf{c}_2
T(\mathbf{c}_1,\mathbf{c}_2)\chi_H(\mathbf{c}_1)\chi_H(\mathbf{c}_2),\\
p\zeta_T&=&-\frac{\gamma}{2}\int\!\!d\mathbf{c}_1\!\!\int\!\!d\mathbf{c}_2
\left(\frac{2c_1^2}{d}-1\right)T(\mathbf{c}_1,\mathbf{c}_2)
\chi_H(\mathbf{c}_1)\chi_H(\mathbf{c}_2), 
\end{eqnarray}
we obtain
\begin{eqnarray}
a^s_{1,1}=-\left[1-\frac{z}{2(1+z)}\right]^2
+\left(1+\frac{2}{z}\right)\frac{z}{1+z}
\left[1-\frac{z}{2(1+z)}\right]\nonumber\\
+\frac{b(p)}{2p\zeta _n}\frac{z^2}{(1+z)^2},
\end{eqnarray}
\begin{eqnarray}
a^s_{1,2}=\frac{1}{p\zeta_T +4p\zeta_n }\left[\left[\frac{z}{2(1+z)}-1\right]
\frac{p(2\zeta_n +\zeta_T )}{(1+z)}\right.\nonumber\\
\left.+p(\zeta_n +\zeta_T )\frac{z}{2(1+z)^2}
-b(p)\frac{z}{(1+z)^2}\right],
\end{eqnarray}
\begin{eqnarray}
a^s_{2,2}=\frac{1}{2p(\zeta_T +3\zeta_n )}
\left[-\frac{p(3\zeta_n +2\zeta_T )}{2(1+z)^2}+\frac{b(p)}{(1+z)^2}\right],\\
a^s_{3i,3j}=\delta_{ij}\frac{c(p)}{2(\zeta_T -\zeta_n )},
\end{eqnarray}
where 
\begin{eqnarray}
b(p)&=&\gamma\int\!\!d\mathbf{c}_1\!\!\int\!\!d\mathbf{c}_2
\frac{c_1^2c_2^2}{d^2}T(\mathbf{c}_1,\mathbf{c}_2)
\chi_H(\mathbf{c}_1)\chi_H(\mathbf{c}_2), \\
c(p)&=&\gamma\int\!\!d\mathbf{c}_1\!\!\int\!\!d\mathbf{c}_2
\chi_H(\mathbf{c}_1)\chi_H(\mathbf{c}_2)\!\!\int\!\!d\boldsymbol{\hat{\sigma}}
\Theta(\mathbf{c}_{12}\cdot\boldsymbol{\hat{\sigma}})
(\mathbf{c}_{12}\cdot\boldsymbol{\hat{\sigma}})c_{1x}c_{2x}.
\end{eqnarray}
These two functions have been evaluated in first Sonine order with the result
\begin{eqnarray}
b(p)&=&-\frac{16(-1+4d(d+1))p+a_2[256-255p+4d(-64+(71+7d)p)]}
{128\sqrt{2}d^2\Gamma(d/2)}\pi^{(d-1)/2}\gamma,\nonumber\\
\end{eqnarray}
\begin{equation}
c(p)=\frac{(-16+5a2)}
{32 \sqrt{2}d\Gamma(d/2)}\pi^{(d-1)/2}\gamma.
\end{equation}

\section{Expressions for $a_{\beta_1,\beta_2}(0)$}
\label{appendixB}
In this  Appendix we evaluate the coefficient $a_{\beta_1,\beta_2}(0)$ for
the specific case in which we have
\begin{eqnarray}
\langle\delta N^2(0)\rangle=0,&\quad&
\langle\delta P_i(0)\delta P_j(0)\rangle=0,\\
\langle\delta E^2(0)\rangle=0,&\quad&\langle\delta N(0)\delta E(0)\rangle=0.
\end{eqnarray}
Taking these relations into account, it is straightforward to obtain
\begin{eqnarray}
\int\!\!d\mathbf{c}_1\!\!\int\!\!d\mathbf{c}_2
\phi_{H}(0,\mathbf{c}_1,\mathbf{c}_2)
&=&-1,\\ 
\int\!\!d\mathbf{c}_1\!\!\int\!\!d\mathbf{c}_2c_{ii}c_{2j}
\phi_{H}(0,\mathbf{c}_1,\mathbf{c}_2)&=&-\frac{1}{2}\delta_{ij},\\
\int\!\!d\mathbf{c}_1\!\!\int\!\!d\mathbf{c}_2c_2^2
\phi_{H}(0,\mathbf{c}_1,\mathbf{c}_2)
&=&-\frac{d}{2},\\
\int\!\!d\mathbf{c}_1\!\!\int\!\!d\mathbf{c}_2 c_1^2c_2^2
\phi_{H}(0,\mathbf{c}_1,\mathbf{c}_2)
&=&-\frac{d(d+2)}{4}(1+a_2).
\end{eqnarray}
With these expressions and the definition of $\bar{\xi}_i$
we get
\begin{eqnarray}
a_{1,1}(0)&=&\frac{1}{(1+z)^2}
\left[\frac{z}{2}(z+2)-\frac{1}{4}(z+2)^2-\frac{d+2}{4d}z^2(1+a_2)\right],\\
a_{1,2}(0)&=&\frac{1}{2(1+z)^2}
\left[-\left(\frac{z}{2}+2\right)+\frac{d+2}{2d}z(1+a_2)\right],\\
a_{2,2}(0)&=&-\frac{1}{(1+z)^2}\left[\frac{3}{4}+\frac{d+2}{4d}(1+a_2)\right],\\
a_{3i,3i}(0)&=&-\frac{1}{2}.
\end{eqnarray}

\section{Langevin equations for the global magnitudes}\label{Sec. 7}

In this Appendix, we will show that it is possible to find a Langevin
description for the fluctuations of the global magnitudes of the system. The
idea is to assume that the global magnitudes obey some equations that can be
decomposed in a ``deterministic part'', which is identified with the
macroscopic equations for a linear perturbation of the HDS, plus a Gaussian
white noise. Because of formulas (\ref{prediction1})-(\ref{prediction3}) let
us study the equations for the magnitudes
\begin{eqnarray}
\delta\tilde{N}(\tau)=\frac{\delta N(\tau)}{N_H^{1/2}(\tau)},\qquad
\delta\tilde{P}_i(\tau)=\frac{\delta P_i(\tau)}{N_H^{1/2}(\tau)v_H(\tau)},
\nonumber\\
\delta\tilde{E}(\tau)=\frac{4\delta E(\tau)}{dmN_H^{1/2}(\tau)v_H^2(\tau)}, 
\end{eqnarray}
in order to deal with processes with time independent variances. 

Let us start with the easiest one, the equation for $\delta\tilde{P}_i(\tau)$.
We can define the function
\begin{equation}
\omega_{i,\mathbf{k}=\mathbf{0}}(\tau)=\frac{1}{n_H(t)v_H(t)}\int\!\! 
d\mathbf{r}\!\!\int\!\!d\mathbf{v}v_i\delta f(\mathbf{r},\mathbf{v},t),
\end{equation}
where $\delta f(\mathbf{r},\mathbf{v},t)\equiv 
f(\mathbf{r},\mathbf{v},t)-f_H(\mathbf{v},t)$, with 
$f_H(\mathbf{v},t)$ the distribution function in the HDS. Then, if the
generic distribution function $f(\mathbf{r},\mathbf{v},t)$ is close enough 
to the HDS one, the linear equation for  
$\omega_{i,\mathbf{k}=\mathbf{0}}(\tau)$ is (see the companion paper)
\begin{equation}
\left(\frac{\partial}{\partial \tau}-p\zeta_T\right)\omega_{i,\mathbf{k}
=\mathbf{0}}(\tau)=0.
\end{equation}
Then it is straightforward 
to see that the equation for the macroscopic deviation 
$\delta\tilde{P}^M$ would be 
\begin{equation}
\left[\frac{\partial}{\partial\tau}+p(\zeta_n-\zeta_T)\right]
\delta\tilde{P}^M(\tau)=0,
\end{equation}
where the superscript ``$M$'' denotes macroscopic. Now, let us suppose that 
the equation for the fluctuating $\delta\tilde{P}$ is of the form
\begin{equation}\label{vii.5}
\left[\frac{\partial}{\partial\tau}+p(\zeta_n-\zeta_T)\right]\delta\tilde{P}_i(\tau)=R_p(\tau),
\end{equation}
with $R_p(\tau)$ a Gaussian white noise with the following properties 
\begin{equation}
\langle R_p(\tau)\rangle_H=0,\qquad
\langle R_p(\tau)R_p(\tau^\prime)\rangle_H=\Gamma_p\delta(\tau-\tau^\prime).
\end{equation}
That is, we consider that the equation describing the dynamics of the
fluctuations can be obtained from the macroscopic equation describing the
evolution of the system. Under these hypothesis we can calculate 
$\langle\delta\tilde{P}(\tau)\delta\tilde{P}(\tau^\prime)\rangle$. 
The solution of equation (\ref{vii.5}) in the long time limit is
\begin{equation}
\delta\tilde{P}_i(\tau)=\int_0^\tau d\tau^\prime e^{-p(\zeta_n-\zeta_T)(\tau-\tau^\prime)}R_p(\tau^\prime),
\end{equation}
and the autocorrelation function is, for $\tau>\tau^\prime$
\begin{equation}
\langle\delta\tilde{P}(\tau)\delta\tilde{P}(\tau^\prime)\rangle
=\frac{\Gamma_p e^{-p(\zeta_n-\zeta_T)(\tau-\tau^\prime)}}{2p(\zeta_n-\zeta_T)}
\left(1-e^{-2p(\zeta_n-\zeta_T)\tau^\prime}\right).
\end{equation}
We are interested in the limit, 
$\tau^\prime\to\infty,\tau\to\infty,\tau^\prime-\tau\sim finite$. 
In this limit we obtain 
\begin{equation}\label{vii.9}
\langle\delta\tilde{P}(\tau)\delta\tilde{P}(\tau^\prime)\rangle_H
=\frac{\Gamma_p}{2p(\zeta_n-\zeta_T)}e^{-p(\zeta_n-\zeta_T)(\tau-\tau^\prime)}.
\end{equation}
Now one can relate this result to the one obtained in the previous
section, equation (\ref{vi.18}), that can be expressed in our variables as
%\begin{equation}
%\langle\delta P_i(\tau)\delta P_i(\tau^\prime)\rangle_H=N_H(\tau)v_H(\tau)v_H(\tau^\prime)\left[\frac{1}{2}+a_{3i3i}\right]e^{p\zeta_T(\tau-\tau^\prime)},
%\end{equation}
%or in our variables
\begin{equation}\label{vii.11}
\langle\delta\tilde{P}(\tau)\delta\tilde{P}(\tau^\prime)\rangle_H=\left[\frac{1}{2}+a^s_{3i,3i}\right]e^{-p(\zeta_n-\zeta_T)(\tau-\tau^\prime)}.
\end{equation}
Comparing equations (\ref{vii.9}) and  (\ref{vii.11}) it is seen that if
\begin{equation}
\Gamma_p=2p(\zeta_n-\zeta_T)\left[\frac{1}{2}+a^s_{3i,3i}\right],
\end{equation}
the Langevin equation (\ref{vii.5}) is in agreement with the results
obtained in the previous section.

Now we will sketch the derivation of the Langevin equations for the
other fluctuating quantities. First of all, we are going to start from the 
macroscopic equation for $\rho_\mathbf{0}$ and  $\varepsilon_\mathbf{0}$ 
defined as
\begin{eqnarray}
\rho_\mathbf{0}\equiv\rho_{\mathbf{k}=\mathbf{0}}(\tau)
&=&\frac{1}{n_H(t)}\int\!\!d\mathbf{r}\!\!\int\!\!d\mathbf{v}\delta
f(\mathbf{r},\mathbf{v},t),\\
\varepsilon_\mathbf{0}\equiv\varepsilon_{\mathbf{k}
=\mathbf{0}}(\tau)&=&\frac{2}{dn_H(t)T_H(t)}\int\!\!d\mathbf{r}\!\!
\int\!\!d\mathbf{v}\frac{1}{2}mv^2\delta f(\mathbf{r},\mathbf{v},t).
\end{eqnarray}
These equations are
\begin{eqnarray}\label{vii.20}
\frac{\partial}{\partial\tau}\rho_\mathbf{0}(\tau)
+p\zeta_n[\rho_\mathbf{0}(\tau)+\varepsilon_\mathbf{0}(\tau)]&=&0,\\
\frac{\partial}{\partial\tau}\varepsilon_\mathbf{0}(\tau)
+p(\zeta_n+\zeta_T)[\rho_\mathbf{0}(\tau)+\varepsilon_\mathbf{0}(\tau)]&=&0,
\end{eqnarray}
from which we can write the equations for 
$\delta\tilde{N}^M=V^{-1/2}n_H^{1/2}\rho_\mathbf{0}$ and
$\delta\tilde{E}^M=V^{-1/2}n_H^{1/2}\varepsilon_\mathbf{0}$ 
\begin{eqnarray}
\frac{\partial}{\partial\tau}\delta\tilde{N}^M&=&-2p\zeta_n\delta\tilde{N}^M
-p\zeta_n\delta\tilde{E}^M,\\
\frac{\partial}{\partial\tau}\delta\tilde{E}^M&=&-p(\zeta_n+\zeta_T)
\delta\tilde{N}^M
-p(2\zeta_n+\zeta_T)\delta\tilde{E}^M.
\end{eqnarray}
As we obtain a linear system of coupled equations, it is convenient to
introduce some new variables to diagonalize the problem:
\begin{eqnarray}
X_1^M&=&\frac{\zeta_n+\zeta_T}{\zeta_n}\delta\tilde{N}^M-\delta\tilde{E}^M,\\
X_2^M&=&\delta\tilde{N}^M+\delta\tilde{E}^M.
\end{eqnarray}
We obtain	
\begin{eqnarray}
\frac{\partial}{\partial\tau}X_1^M&=&-p\zeta_nX_1^M,\label{vii.30}\\
\frac{\partial}{\partial\tau}X_2^M&=&-p(3\zeta_n+\zeta_T)X_2^M.\label{vii.31}
\end{eqnarray}
Let us suppose now that the equations for $X_i(\tau)$ have the form of the
macroscopic equations (\ref{vii.30}) and (\ref{vii.31}) plus a Gaussian
random noise which variance has to be computed to reproduce the
results that we have obtained for the correlation function of the
fluctuations of $N$ and $E$. The equations for the fluctuating
variables $X_1$ and $X_2$ are thus
\begin{eqnarray}
\left[\frac{\partial}{\partial\tau}+p\zeta_n\right]X_1&=&R_1(\tau), \\
\left[\frac{\partial}{\partial\tau}+p(3\zeta_n+\zeta_T)\right]X_2&=&R_2(\tau).
\end{eqnarray}
Then, if we suppose that the noise terms have zero mean and its correlation 
function is delta-correlated in time
\begin{equation}
\langle R_i(\tau)\rangle_H=0,\quad
\langle R_i(\tau^\prime)R_j(\tau)\rangle_H=\Gamma_{ij}\delta(\tau^\prime-\tau),
\end{equation}
we can obtain the values of the amplitudes of the noise term $\Gamma_{ij}$
in the same way we have done with the momentum, that is comparing with the
results we have obtained in the previous section. We obtain
\begin{eqnarray}
\Gamma_{11}&=&-8p\zeta_n\frac{(1+z)^2}{z^2}A_{11},\\
\Gamma_{22}&=&-8p(3\zeta_n+\zeta_T)(1+z)^2A_{22},\\
\Gamma_{12}&=&-4p(4\zeta_n+\zeta_T)\frac{(1+z)^2}{z}A_{12}.
\end{eqnarray}

These calculations show that it is possible to describe the dynamics
of fluctuations in the HDS in terms  
of some Langevin equations with Gaussian white noises. Due to the Gaussian
nature of the noises and given that the equations are linear, the
probability distribution function for those processes 
will be also Gaussian in agreement with our simulations. 
It is worth pointing out 
that, although the amplitude of the noises are known, they are not
related in a simple way 
to the cooling rates $\zeta_n$ and $\zeta_T$ that appear in the
``deterministic part'' of the equations.


\begin{thebibliography}{10}

\bibitem{companion}
M.~I. {Garc\'{\i}a de Soria} {\it et~al.},   (2008).

\bibitem{bl88}
M. Bramson and J. Lebowitz, Phys. Rev. Lett. {\bf 61},  2397  (1988).

\bibitem{OZ78}
A. Ovchinikov and Y. Zeldovitch, Chem. Phys. {\bf 28},  214  (1978).

\bibitem{tw83}
D. Toussaint and F. Wilczek, J. Chem. Phys. {\bf 78},  2642  (1983).

\bibitem{rk83}
S. Redner and K. Kang, Phys. Rev. Lett. {\bf 51},  1729  (1983).

\bibitem{l03}
F. Leyvraz, Phys. Rep. {\bf 383},  95  (2003).

\bibitem{cdt04s}
F. Coppex, M. Droz, and E. Trizac, Phys. Rev. E {\bf 69},  011303  (2004).

\bibitem{ks01}
P. Krapivsky and C. Sire, Phys. Rev. Lett. {\bf 86},  2494  (2001).

\bibitem{t02}
E. Trizac, Phys. Rev. Lett. {\bf 88},  160601  (2002).

\bibitem{ptd02}
J. Piasecki, E. Trizac, and M. Droz, Phys. Rev. E {\bf 66},  066111  (2002).

\bibitem{llf06}
A. Lipowski, D. Lipowska, and A. Feirrera, Phys. Rev. E {\bf 73},  032102
  (2006).

\bibitem{cdt04}
F. Coppex, M. Droz, and E. Trizac, Phys. Rev. E {\bf 70},  061102  (2004).

\bibitem{nebo97}
T.~P.~C. van Noije, M.~H. Ernst, R. Brito, and J.~A.~G. Orza, Phys. Rev. Lett.
  {\bf 79},  411  (1997).

\bibitem{bmr98}
J.~J. Brey, F. Moreno, and M.~J. Ruiz-Montero, Phys. Fluids {\bf 10},  2965
  (1998).

\bibitem{bgmr05}
J.~J. Brey, M.~I. {Garc\'{\i}a de Soria}, P. Maynar, and M.~J. Ruiz-Montero,
  Phys. Rev. Lett. {\bf 94},  098001  (2005).

\bibitem{bdgm06}
J.~J. Brey, A. Dom\'inguez, M.~I. {Garc\'{\i}a de Soria}, and P. Maynar, Phys.
  Rev. Lett. {\bf 96},  158002  (2006).

\bibitem{ec81}
M.~H. Ernst and E.~G.~D. Cohen, J. Stat. Phys. {\bf 25},  153  (1981).

\bibitem{bgmr04}
J.~J. Brey, M.~I. {Garc\'{\i}a de Soria}, P. Maynar, and M.~J. Ruiz-Montero,
  Phys. Rev. E {\bf 70},  011302  (2004).

\bibitem{allen}
M.~P. Allen and D.~J. Tildesley, {\em Computer Simulation of Liquids} (Oxford
  Science Publications, Bristol, 1987).

\bibitem{bird}
G.~A. Bird, {\em Molecular Gas Dynamics and the Direct Simulation of Gas Flows}
  (Clarendon, Oxford, 1994).

\bibitem{chandler}
D. Chandler, {\em Introduction to Modern Statistical Mechanics} (Oxford
  University Press, New York, 1987).

\end{thebibliography}
\end{document}